\documentclass[apj, numberedappendix]{emulateapj}  

\newcommand*{\unit}[1]{\ensuremath{\mathrm{\, #1}}}

\newcommand*{\degree}{\ensuremath{^{\circ}}}
\newcommand*{\km}{\unit{km}}
\newcommand*{\ps}{\unit{s}^{-1}}
\newcommand*{\pcmc}{\unit{cm}^{-3}}	
\def \hsi {{\it RHESSI\ }}	
\def \goes {{\it GOES\ }}
\def \soho {{\it SOHO}}

\begin{document}


\title{Double Coronal Hard and Soft X-ray Source Observed by {\it RHESSI}: Evidence for 
Magnetic Reconnection and Particle Acceleration in Solar Flares}

\author{Wei Liu\altaffilmark{1}$^,$\altaffilmark{2}, 
 Vah\'{e} Petrosian\altaffilmark{2}, Brian R. Dennis\altaffilmark{1}, and Yan Wei Jiang\altaffilmark{2}}


\altaffiltext{1}{Solar Physics Laboratory (Code 671), Heliophysics Science Division, 
 NASA Goddard Space Flight Center, Greenbelt, Maryland 20771; weiliu@helio.gsfc.nasa.gov}
\altaffiltext{2}{Center for Space Science and Astrophysics, Department of Physics, Stanford University, 
  Stanford, California 94305}	

\journalinfo{To appear in ApJ March 20, 2008; online at astroph http://arxiv.org/abs/0709.1963}
\submitted{Received 2007 September 12; accepted 2007 December 06}

\begin{abstract}

We present data analysis and interpretation of an M1.4-class flare observed with
the Reuven Ramaty High Energy Solar Spectroscopic Imager ({\it RHESSI}) on April 30, 2002. 
This event, with its footpoints occulted by the solar limb, exhibits a rarely observed, 
but theoretically expected, double-source structure in the corona. 
The two coronal sources, observed over the 6--30~keV range, appear at different altitudes and
show energy-dependent structures with the higher-energy emission being closer together.
 Spectral analysis implies that the emission at higher energies in the inner region between the two sources
is mainly {\it nonthermal}, while the emission at lower energies in the outer region is primarily {\it thermal}.
The two sources are both visible for about 12 minutes and have similar light curves and power-law
spectra above about 20~keV.
These observations suggest that the		
magnetic reconnection site lies	 
between the two sources.  Bi-directional outflows of the released energy in the form of
turbulence and/or particles	
from the reconnection site can be the source of the observed radiation. 
The spatially resolved thermal emission below about 15~keV, on the other hand, 
indicates that the lower source has a larger emission measure
but a lower temperature than the upper source. 
This is likely the result of the	
differences in the magnetic field and plasma density
of the two sources.
	


\end{abstract}



\section{Introduction}
\label{chp2LT_intro}

\begin{figure}[thbp]      
 \epsscale{0.9}   
 \plotone{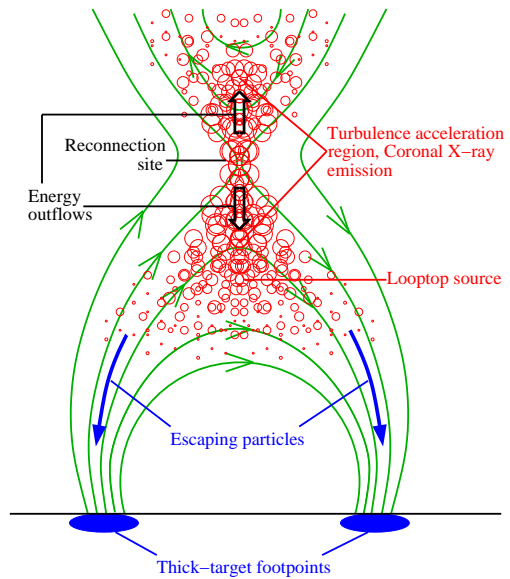}
 \caption[Cartoon_model]
 {Sketch of the stochastic acceleration model \citep{HamiltonR1992ApJ...398..350H, ParkB1995ApJ...446..699P,
 PetrosianV2004ApJ...610..550P} proposed for	
 solar flares. 
 The green curves are magnetic field lines in a possible configuration; the red circles 
 represent turbulence or plasma waves that are generated during magnetic reconnection.
 } \label{fig_model}
 \end{figure}

Magnetic reconnection is believed to be the main energy release mechanism in solar flares. 
In the classical 	
reconnection model \citep[e.g.,][]{PetschekH1964psf..conf..425P},
magnetic field annihilation in a current sheet 
produces outflows of high speed plasma	
in opposite directions (see Figure~\ref{fig_model}).  This process can generate turbulence		
that accelerates particles and heats the background plasma stochastically 
\citep[e.g.,][]{RamatyR1979AIPC...56..135R, HamiltonR1992ApJ...398..350H, ParkB1995ApJ...446..699P,
MillerJ1996ApJ...461..445M, PetrosianV2004ApJ...610..550P}.
Radio emission and hard and soft X-rays (HXRs and SXRs) 
produced by the high-energy particles and hot plasma 
are expected to show signatures of	
the two oppositely directed outflows.	
Specifically, one would expect to see two distinct X-ray sources, 	
one above and one below the reconnection region (in the case of a vertical current sheet).

It is well established that many flares have SXR and HXR emission arising both
from the source near the top of the loop \citep[loop-top source, e.g.,][]
{MasudaS1994PhDT.......240M, PetrosianV2002ApJ...569..459P, LiuW2004ApJ...611L..53L, 
JiangY2006ApJ...638.1140J, LiuW2006PhDT........35L, BattagliaM_LTFP_2006A&A...456..751B} 
and from a pair of footpoint sources \citep[e.g.,][]{HoyngP1981ApJ246L.155.HXRfp, SakaoT1994PhDT.......335S,
SuiL2002SoPh..210..245S, Saint-Hilaire2007SolPh.FPasym}.
The loop-top source is believed to be near the reconnection site and produced by freshly 
accelerated particles and/or heated plasma.
Observations of the expected two distinct X-ray sources 
above and below the reconnection region have been rarely reported. This may be due to limited sensitivity, 
dynamic range, and/or spatial resolution of the instruments, because one source may be much dimmer than the other,
the two sources may be too close together to be resolved,	
or both may be much weaker than the footpoints.

\citet{SuiL2003ApJ...596L.251S} and \citet{SuiL2004ApJ...612..546S}   
reported a second coronal source that appeared above a stronger
loop-top source in the 2002 April 15 flare and  
in another two homologous flares.  
They suggested that there was a current sheet existing between the two sources. 
Recently, in one of the events reported by \citet{SuiL2004ApJ...612..546S}, 
\citet{WangT.SuiL2007ApJ}	
discovered high speed outflows revealed by Doppler shifts measured by
the Solar Ultraviolet Measurements of Emitted Radiation (SUMER) instrument on board 
the Solar and Heliospheric Observatory (\soho).  This provides more evidence of magnetic 
reconnection.	
\citet{VeronigA2006A&A...446..675V} 
also found a second coronal source appearing briefly
in the 2003 November 03 X3.9 flare \citep{LiuW2004ApJ...611L..53L, Dauphin2006A&A}. 
 \citet{LiYPGanWQ2007AdSpR}	
reported another \hsi flare, occurring on 2002 November 02, that shows a similar double coronal source morphology.
They interpreted the two sources as thermal emission because no HXR emission was detected
above 25~keV and the footpoints were occulted. In their event, however, the two sources have somewhat
different temporal evolution with the flux of the upper source peaking about 18 minutes later than
that of the lower source. 
In radio wavelengths, \citet{PickM2005ApJ}	
reported a double-source structure observed in the 2002 June 02 flare with the 
multi-frequency Nan\c{c}ay Radio-heliograph (432--150 MHz).  Due to its low brightness and/or 
technical difficulties, X-ray imaging spectroscopy of the weaker coronal source was not 
available or has not been studied for the above \hsi events \citep{SuiL2003ApJ...596L.251S,
SuiL2004ApJ...612..546S, VeronigA2006A&A...446..675V, LiYPGanWQ2007AdSpR}.

We report here another flare with two distinct coronal X-ray sources that occurred on April 30, 2002.
The brightness of the upper source relative to the lower source
is larger and the upper source stays longer ($\sim$12~minutes) than those (3--5~minutes) 	
of \citet{SuiL2004ApJ...612..546S}.  In addition, the footpoints are occulted
by the solar limb, and thus the spectra of the coronal sources are not contaminated by the footpoints
at high energies.  
This makes a stronger case for the double coronal source phenomenon and allows for more detailed analysis, 
including X-ray imaging spectroscopy and temporal evolution of the individual sources. 	
Analysis of the decay phase of this flare was originally reported by \citet{JiangY2006ApJ...638.1140J} as an example
of suppression of thermal conduction and/or continuous heating attributed to the presence of plasma turbulence.
Here we extend the analysis throughout the whole course of the flare.

Early in the flare, 	
the two coronal sources are close together	
and the source morphology exhibits a double-cusp or ``X" shape, possibly indicating where
magnetic reconnection takes place. As the flare evolves, the two sources gradually separate from each other.
Both sources exhibit energy dependent structure similar to that found for the flares reported by 
\citet{SuiL2004ApJ...612..546S}		
and \citet{LiuW2004ApJ...611L..53L}.  
In general, for the lower source, higher-energy emission comes from higher altitudes, 
while the opposite is true for the upper source.
Imaging spectroscopy shows that the two sources have very similar nonthermal components
and light curves.
These observations suggest that the two HXR/SXR coronal sources are produced by	
intimately related populations of accelerated/heated electrons resulting from energy release
in the same reconnection region, which
most likely lies between the two sources.  These are consistent with the general picture outlined above.
We also find that the two sources have different thermal components, with a lower temperature 
and larger emission measure for the lower source.  Different magnetic topologies 
and plasma densities of the two sources can be the causes of such differences.
 
We present the observations and data analysis in \S\ref{chp2LT_obs} and our physical interpretation
in \S\ref{section_interpret}.  We conclude this paper with a brief summary and discussion in
\S\ref{chp2LT_summary}. Details of specific \hsi spectral analysis techniques are 
provided in Appendix \ref{appendixA}.


\section{Observations and Data Analysis}
\label{chp2LT_obs}

 \begin{figure}[thbp]      
 \epsscale{1.1}  
 \plotone{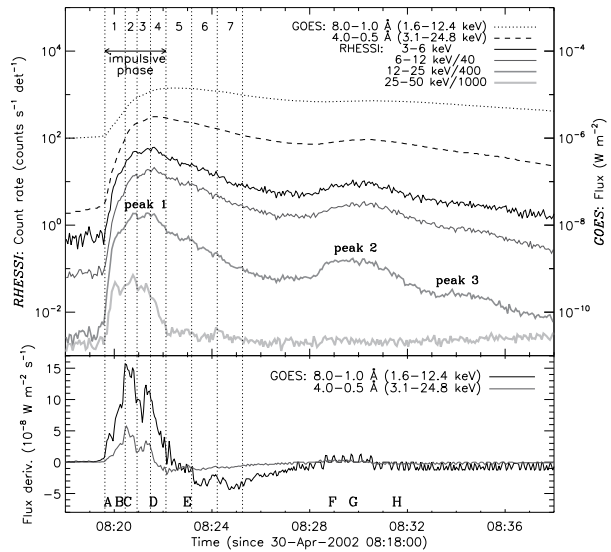}		
 \caption[Flare of 04/30/2002: \hsi light curves]
 { {\it Top}: \hsi and {\it GOES}-10 light curves of the 2002 April 30 M1.4 flare. 
 The \hsi count rates ({\it left scale}) are averaged over every 4 seconds, with scaling factors of 1, 1/40, 1/400, and 1/1000 
 for the energy bands 3--6, 6--12, 12--25, and 25--50 keV, respectively.
 The \goes fluxes ({\it right scale}) in the bandpass of 8--1 {\AA} (1.6--12.4 keV) and 4.0--0.5 {\AA} (3.1--24.8 keV) 
 are in a cadence of 3 seconds. The vertical dotted lines mark the seven time intervals used for imaging
 spectroscopy (see Figures~\ref{chp2LT_imgspec} and \ref{fig_index}).		
 Intervals 1--4 correspond to the impulsive phase according to the 25--50~keV light curve.
 Three peaks on the 12--25~keV curves are also marked.
   {\it Bottom}: Time derivatives of the {\it GOES} fluxes, showing the Neupert-type behavior. 
 The capital letters (A--H) at the bottom mark the central times of the intervals of the images
 shown in Figure~\ref{chp2LT_mosaic}.
 Note that the energy release episode of peak~2 may actually start at $\sim$08:25~UT when
 the exponential decay rate of the \hsi light curves (3--25~keV) decreases
 and the \goes time derivative (8--1~\AA) starts to increase.
 } \label{lc.ps}
 \end{figure}
The event under study, classified as a {\it Geostationary Operational Environment Satellite} ({\it GOES}) 
M1.4 flare, started at 08:19~UT on April 30, 2002.  Figure~\ref{lc.ps} (top panel) shows the \hsi light curves 
in four energy bands between 3 and 50~keV together with the fluxes of the two \goes channels (1--8 and 0.5--4.0 {\AA}).
During the full course of the flare, \hsi is in the A1 attenuator state, i.e. with the thin shutters in. 
According to the 12--25~keV light curve, there are three peaks  
that are progressively weaker and softer.
Above 25~keV there is no detectable count rate increase beyond the first peak that 	
we refer to as the impulsive phase.

The temporal trend of the time derivatives (Figure~\ref{lc.ps}, bottom panel) of the \goes fluxes 
mimics that of the \hsi 25--50~keV count rate.  This type of correlation is known as the Neupert effect 
\citep{NeupertW1968ApJ...153L..59N} and has been observed in many (but not all) flares \citep{DennisB1993SoPh..146..177D,
VeronigA2005ApJ...621..482V, LiuW2006ApJ...649.1124L}.
Such flares are usually observed on the solar disk where HXRs are seen from the footpoints, indicating
prompt energy release and impulsive heating of the chromosphere by the nonthermal electrons.
The hot and dense plasma resulting from the subsequent chromospheric evaporation 
\citep{NeupertW1968ApJ...153L..59N} then fills the loop and gives rise to the gradual SXR increase
\citep[][Chapter 8%
  \footnote{Available at http://sun.stanford.edu/$\sim$weiliu/thesis/wei\_thesis.pdf.}%
]{LiuW2006PhDT........35L}.
In this flare, however, the footpoints are occulted by the limb (as we show below).  Thus, the presence of the Neupert 
effect here implies that the coronal impulsive HXRs are produced by the same nonthermal electrons that 
further propagate down to the footpoints and drive chromospheric evaporation there.

%
 \begin{figure}[thbp]      
 \epsscale{1.2}	  
 \hspace{-0.6cm}
 \plotone{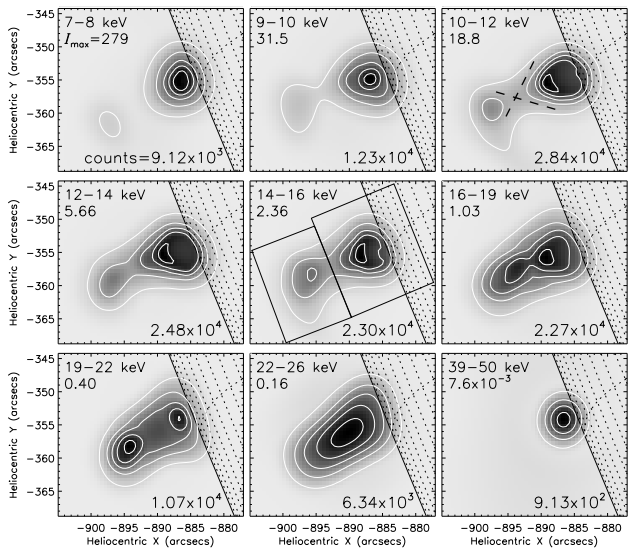}	
 \caption[PIXON images at 08:20:27--08:20:56~UT in different energy bands]
 {\hsi PIXON images in different energy bands at 08:20:27--08:20:56~UT  (interval~2 in Figure~\ref{lc.ps}),  
 around the maximum of the main (first) HXR peak. 
 We used the PIXON background model and detectors 3--6 and 8, which yield a resolution
 of $\sim4 \farcs 6$ determined from the FWHM of the point spread function obtained by simulation.
   Note that the PIXON algorithm, under favorable conditions, 
   can achieve a resolution as small as a fraction \citep[see][\S A8]{AschwandenMJ2004SoPh.HESSI.img.soft} of the 
   FWHM resolution of the finest detector used ($6 \farcs 8$ for detector 3 in our case).
 The contour levels are 10\%, 30\%, 50\%, 70\%, and 90\% of the maximum, $I_\unit{max}$ 
 (shown in the upper left corner of each panel, 
 in units of photons cm$^{-2} \ps \unit{arcsec}^{-2}$), of each individual image.  The numbers in the lower right
 corners are the total counts accumulated by the detectors used for image reconstruction.
 The heliographic grid spacing is 1\degree. The boxes shown in the 14--16 keV panel are used to obtain
 the fluxes and centroids of the two sources in all the images at this time (see text).  
 The two black dashed lines in the 10--12~keV panel forming an ``X" show the possible
 configuration of the reconnecting magnetic field.
 } \label{hsi_multi.ps}
 \end{figure}
 \begin{figure}[thbp]      
 \epsscale{0.57}  
 \plotone{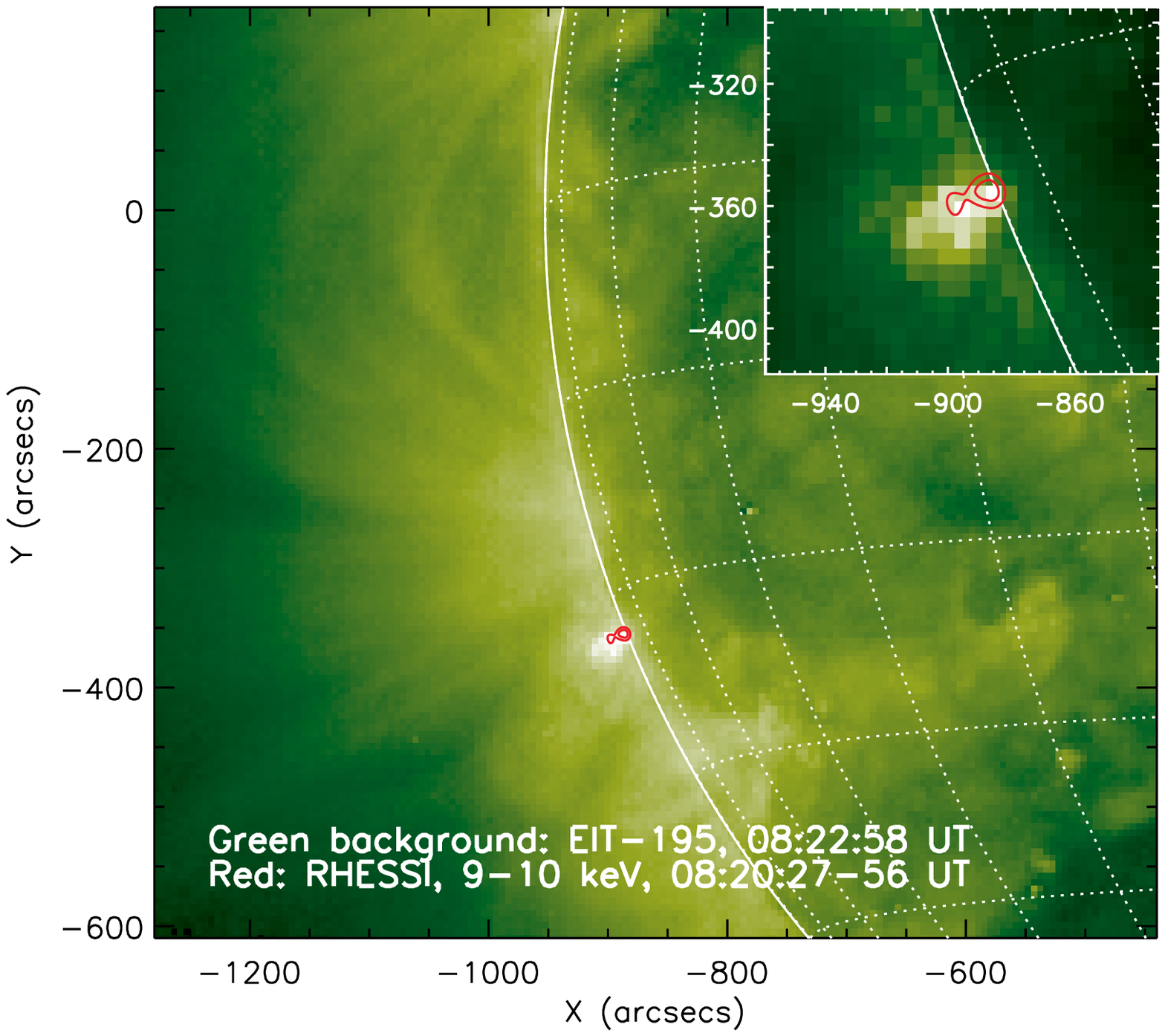}	
 \plotone{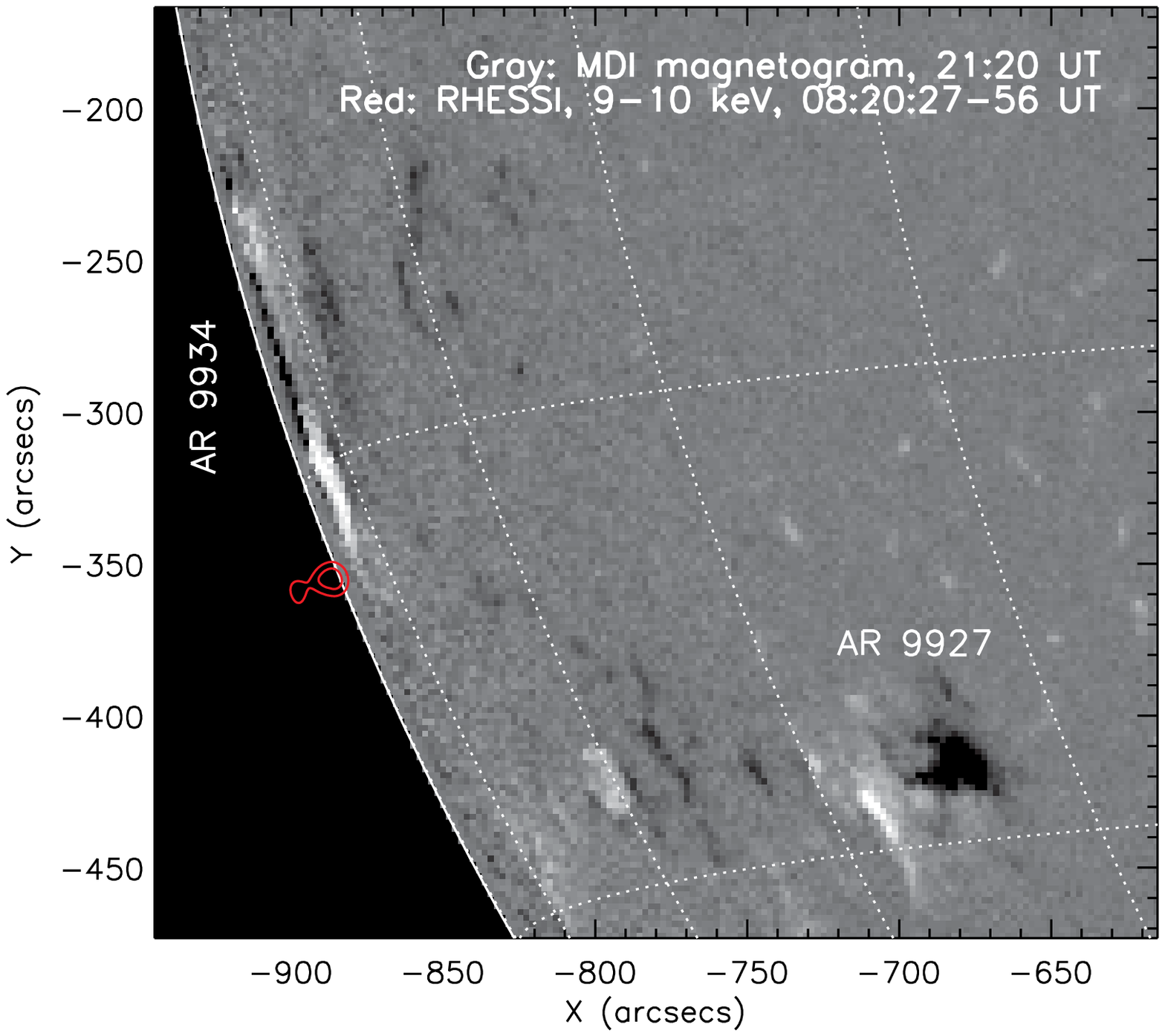}      
 \plotone{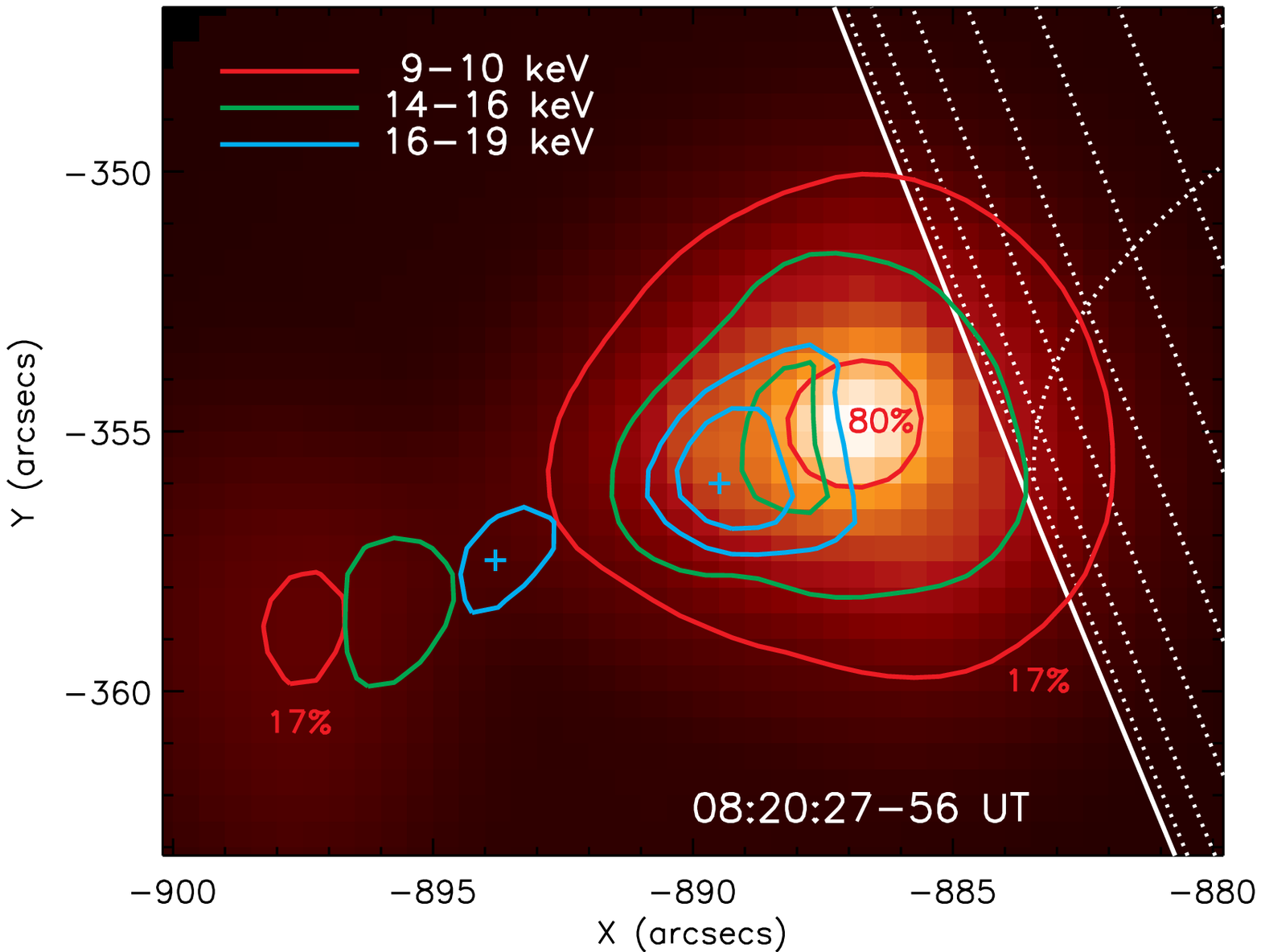}
 \plotone{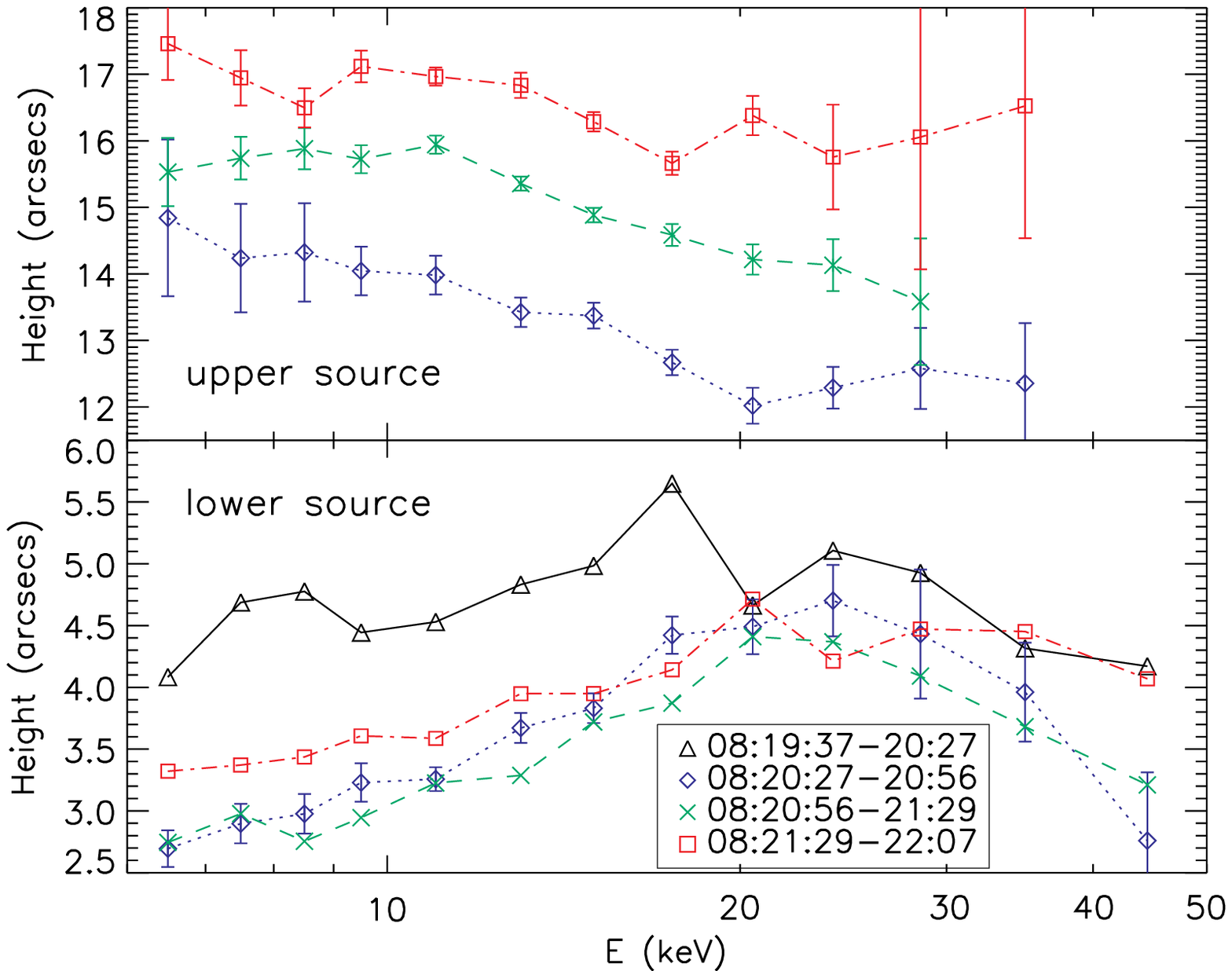}
 \caption[Overlay of images in four energy bands]
 { {\it Upper left}: {\it SOHO}/EIT 195 \AA\ image at 08:22:58~UT in the background, superimposed with 
 \hsi contours in red at 9--10 keV and 08:20:27--08:20:56~UT.  The insert shows the zoomed view of the \hsi source 
 and co-spatial EIT emission (with a different color scale for better contrast). 
   {\it Upper right}: {\it SOHO}/MDI magnetogram taken at 21:20~UT (some 13 hours after the flare), 
 overplotted with the same \hsi 9--10 keV contours in red.  The NOAA active regions (ARs) are labeled.
 The heliographic grid spacing is 10\degree\ in the two upper panels.
   {\it Lower left}: Overlay of images, same as those shown in Figure~\ref{hsi_multi.ps}, in three energy bands 
 as indicated in the legend.
 The contour levels are at 17\% \& 80\% (9--10~keV), 47\% \& 90\% (14--16~keV), and 80\% \& 90\% 
 (16--19~keV) of the maximum brightness of individual images. In each image, two contours
 appear in the lower coronal source, while only the lower-level contour is present in the upper
 source because of its faintness.  The two plus signs mark the centroids (separated by $4\farcs 6 \pm 0\farcs 3$)
 of the lower and upper 16--19~keV sources inside the 90\% and 80\% contours, respectively.
 The heliographic grid spacing is 1\degree.
   {\it Lower right}: Height above the limb of the centroids for the upper and lower coronal sources 
 plotted as a function of energy for time intervals 1--4 marked in Figure~\ref{lc.ps}.
 Note that during the first interval, only one source is detected and is shown as the lower source.
 For clarity, uncertainties are shown for only one time interval for the lower source and they are similar
 at other times.
 } \label{structure}
 \end{figure}

The spatial morphology of the flare is shown in Figure~\ref{hsi_multi.ps} in X-rays and in Figure~\ref{structure}
in EUV.  Figure~\ref{hsi_multi.ps} shows PIXON \citep{MetcalfT1996ApJ...466..585M, HurfordG2002SoPh..210...61H}
 images at different energies integrated over the interval of
08:20:27--08:20:56~UT (marked \#2 in Figure~\ref{lc.ps}) during the first HXR peak.  
 As can be seen, this flare occurred on the east limb,
and the X-ray emission at all energies (even as high as 39--50~keV) appeared above the limb, suggesting that 
the footpoints were occulted.  This conclusion is supported by \soho\ observations shown in Figure~\ref{structure}.  
The top left panel shows an EUV Imaging Telescope (EIT) 
195 \AA\ image taken at 08:22:58~UT (just two minutes after the \hsi images in Figure~\ref{structure}).  
The red contours are for the \hsi image at 9--10~keV shown in Figure~\ref{hsi_multi.ps}.  
The \hsi source is co-spatial with the brightening in the EIT image, which is 
clearly above the limb.  The flare occurred near the region where large-scale trans-equatorial loops are rooted, presumably behind the limb.  
There was no brightening on the disk detected by EIT, nor was an active region seen in the \soho\ 
Michelson Doppler Imager (MDI) magnetograms in the vicinity of this flare.  EIT and MDI have spatial 
resolutions of $2\farcs 6$ and $4\arcsec$, respectively, both better than the 
$6 \farcs 8$ resolution of the finest \hsi detector (\#3) used here.   	
The top right panel of Figure~\ref{structure} shows the MDI magnetogram at 21:20~UT, 
about 13 hours after the flare.  At this time, NOAA AR 9934 had just appeared on the disk 
next to the \hsi source due to the solar rotation.	
This suggests that the flare took place in this active region when it was still behind the limb.
Because of the large size of the active region, it is difficult to determine the possible locations
of the footpoints of the flare and to estimate the approximate altitudes of the coronal sources.

\subsection{Source Structure: Energy Dependence}	
\label{chp2LT_morphovsE}

Let us now return to Figure~\ref{hsi_multi.ps} and examine in detail the energy-dependent morphology of the flare.
At the lowest energy shown (7--8~keV), there are two distinct sources, which we call the lower and upper 
coronal sources. The centroids of both sources are above the solar limb, and the upper source is dimmer.
At a slightly higher energy, 9--10~keV, the sources appear closer together and a cusp shape develops between them. 
This trend is more pronounced at higher energies (10--19~keV) and the two sources (particularly the lower one) 
seem to have a feature convex toward each other, mimicking the ``X" shape of the magnetic field lines in the
standard reconnection model. Meanwhile, the relative brightness of the upper source increases with energy.

The change in source altitude with energy is shown more clearly in the lower left panel of
Figure~\ref{structure}. The upper coronal source shifts toward lower altitudes with increasing energy
while the lower coronal source behaves oppositely.	
At 16--19~keV, 	
the two sources, while being spatially resolved, are closest together with their centroids
separated by $4\farcs 6 \pm 0\farcs 3$ (see Figure~\ref{structure}, {\it lower left}).

We can appreciate this more quantitatively by looking at the heights (above the limb) of the centroids
of the upper and lower coronal sources as a function of energy.  This is shown in the lower right panel of
Figure~\ref{structure}. 
The boxes depicted in the middle panel of Figure~\ref{hsi_multi.ps} were used to obtain the centroid positions.
The error bars were obtained from the centroid position uncertainties in the same images reconstructed 
with the visibility-based forward-fitting algorithm currently available in the \hsi software. 
The energy-dependent pattern is clearly present;	
that is, the centroid of the upper (lower) source shifts to lower (higher) altitudes with increasing energy. 
We note that, at very high energies ($\gtrsim 20$~keV), this pattern becomes obscure 
(see Figures~\ref{hsi_multi.ps} and \ref{structure}), but the uncertainties in the source
locations become large due to low count rates.  

   Three other time intervals during the first HXR peak were also analyzed 
and the results are plotted in Figure~\ref{structure}, exhibiting similar patterns. 
At the very beginning of the flare (08:19:37--08:20:27~UT), only one source is visible and we assign
its centroid ({\it black triangles}) to the lower source since it is the main source.
   As mentioned earlier, the second and third HXR peaks are weaker and softer, which does not allow for this kind of 
detailed analysis with narrow energy bins.  We defer our physical interpretation
of these observations to \S\ref{interp_morphovsE}.



\subsection{Source Structure: Temporal Evolution}	
\label{section_time_01}

We now change our perspective, using relatively wider energy bins as a trade-off for finer time
resolution (compared with the above analysis), and examine the temporal evolution of the source 
structure throughout the full course of the flare.
 \begin{figure}[thbp]      
 \epsscale{0.5}	
 \plotone{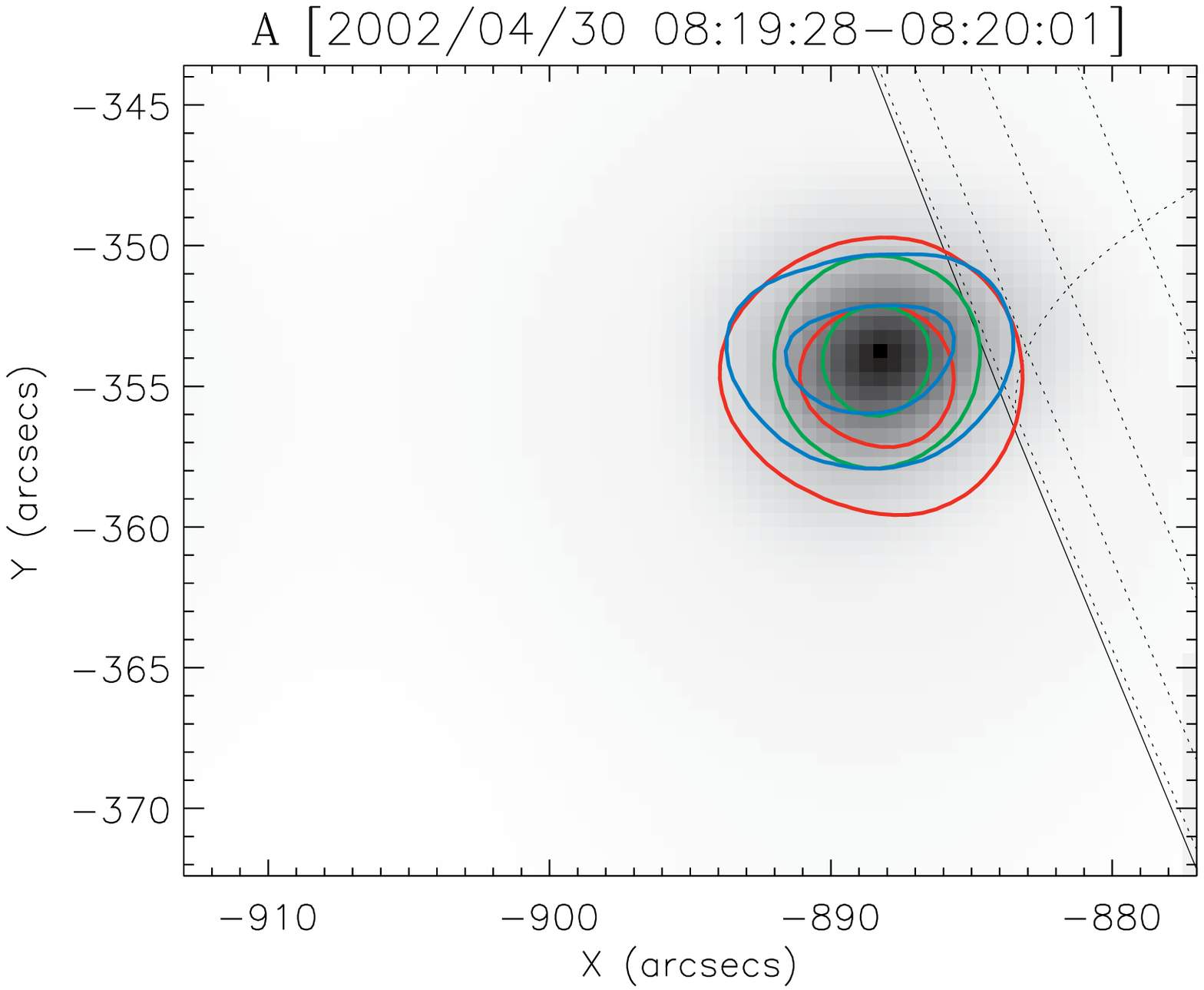}	
 \plotone{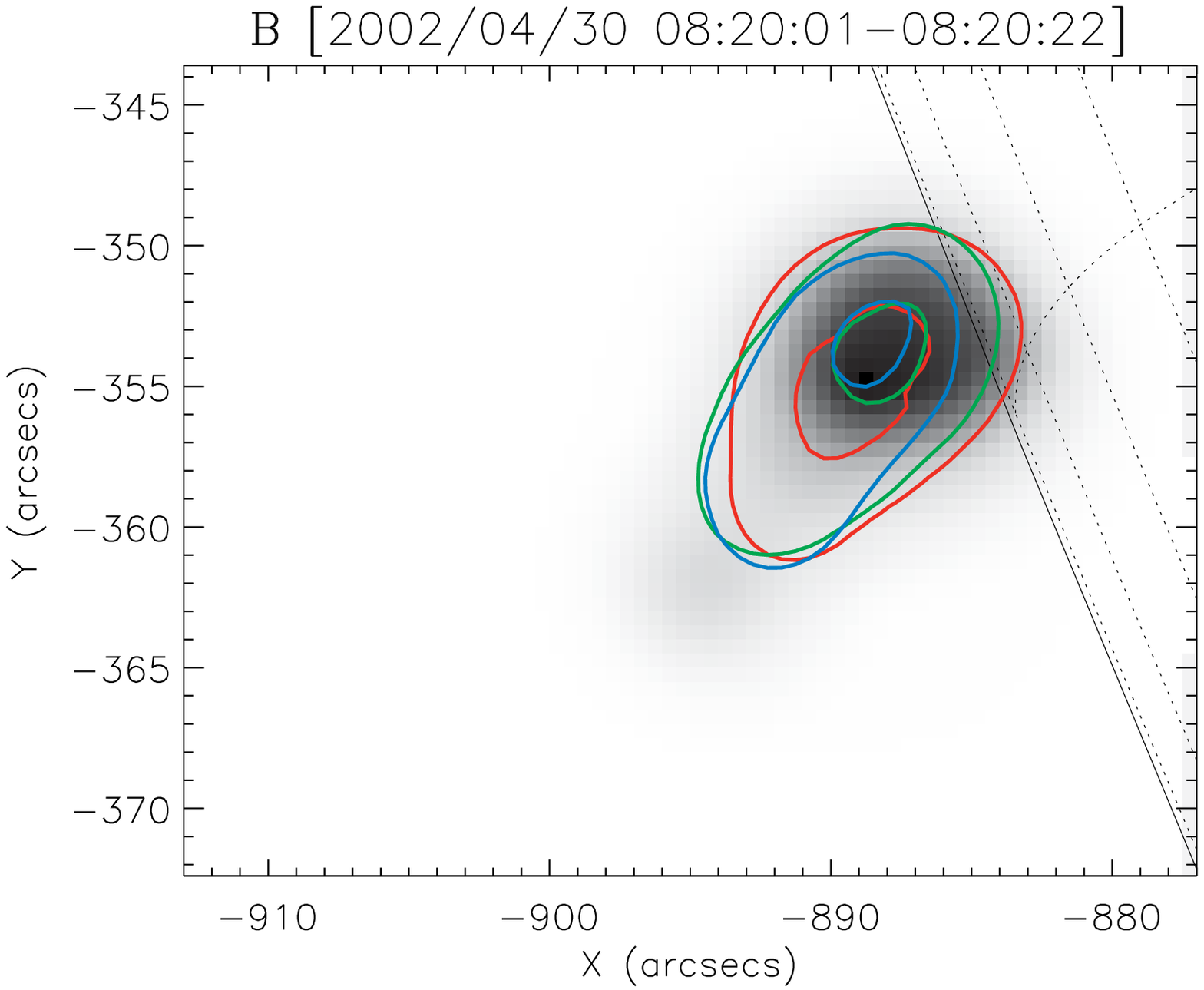}
 \plotone{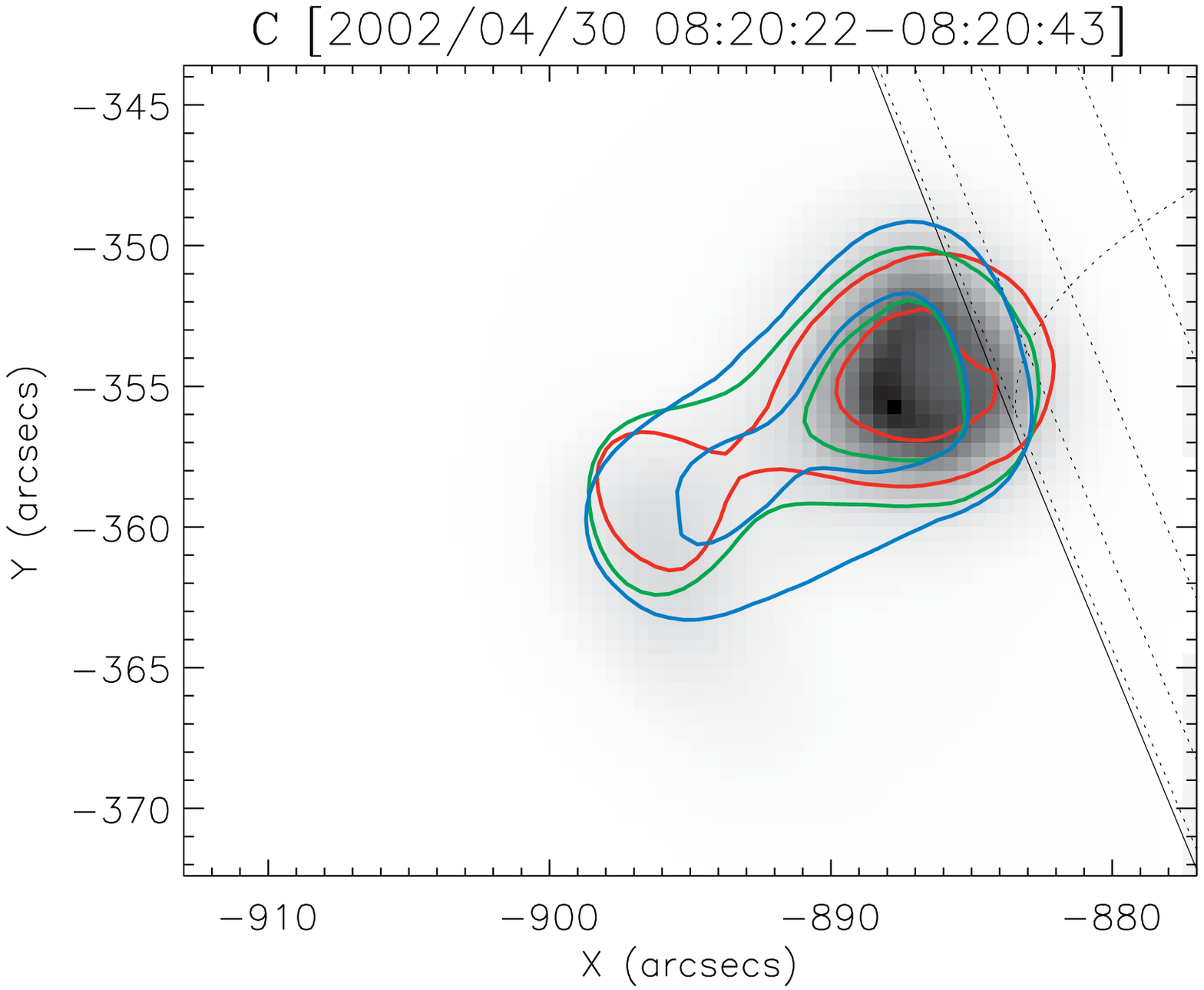}
 \plotone{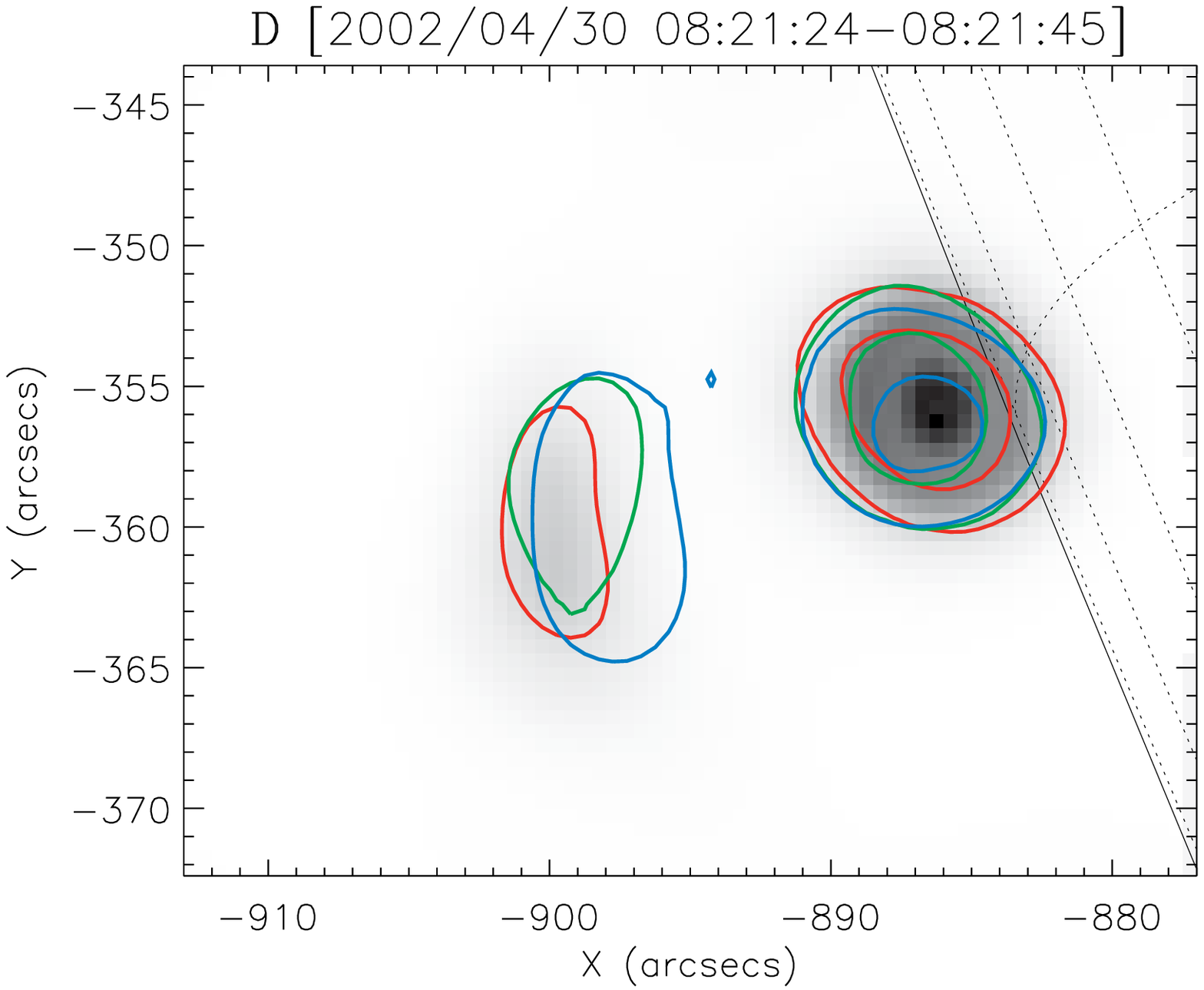}
 \plotone{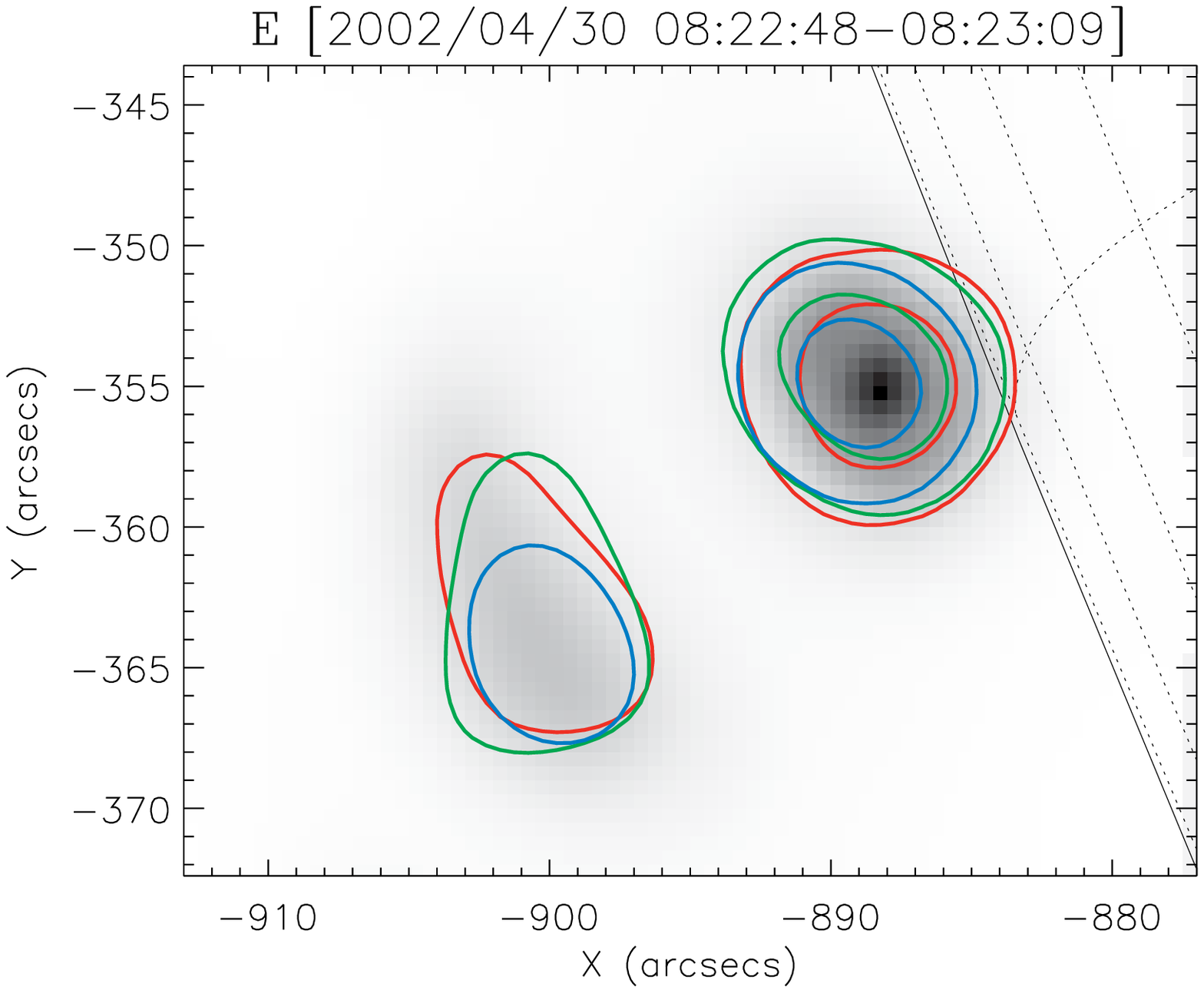}
 \plotone{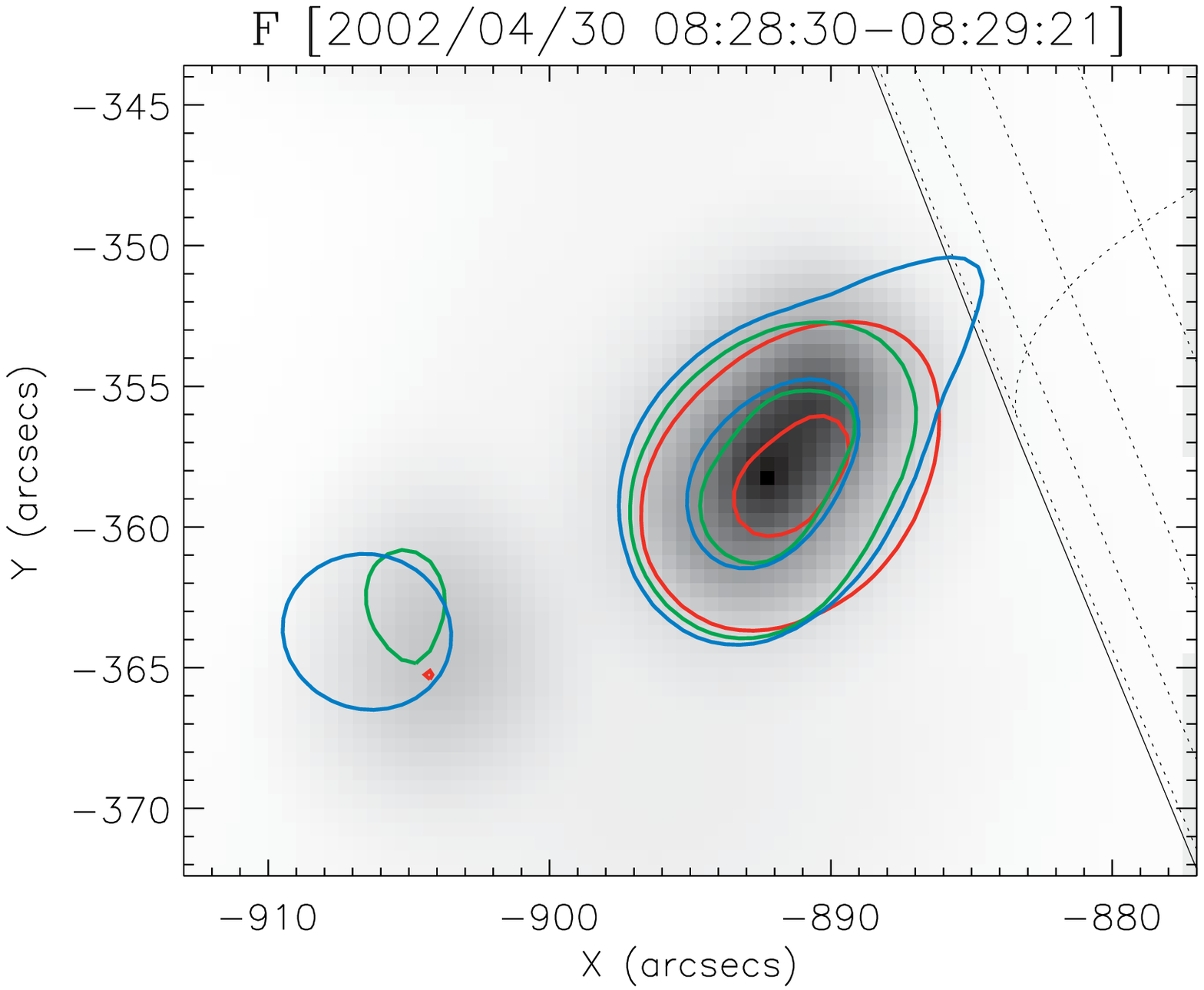}
 \plotone{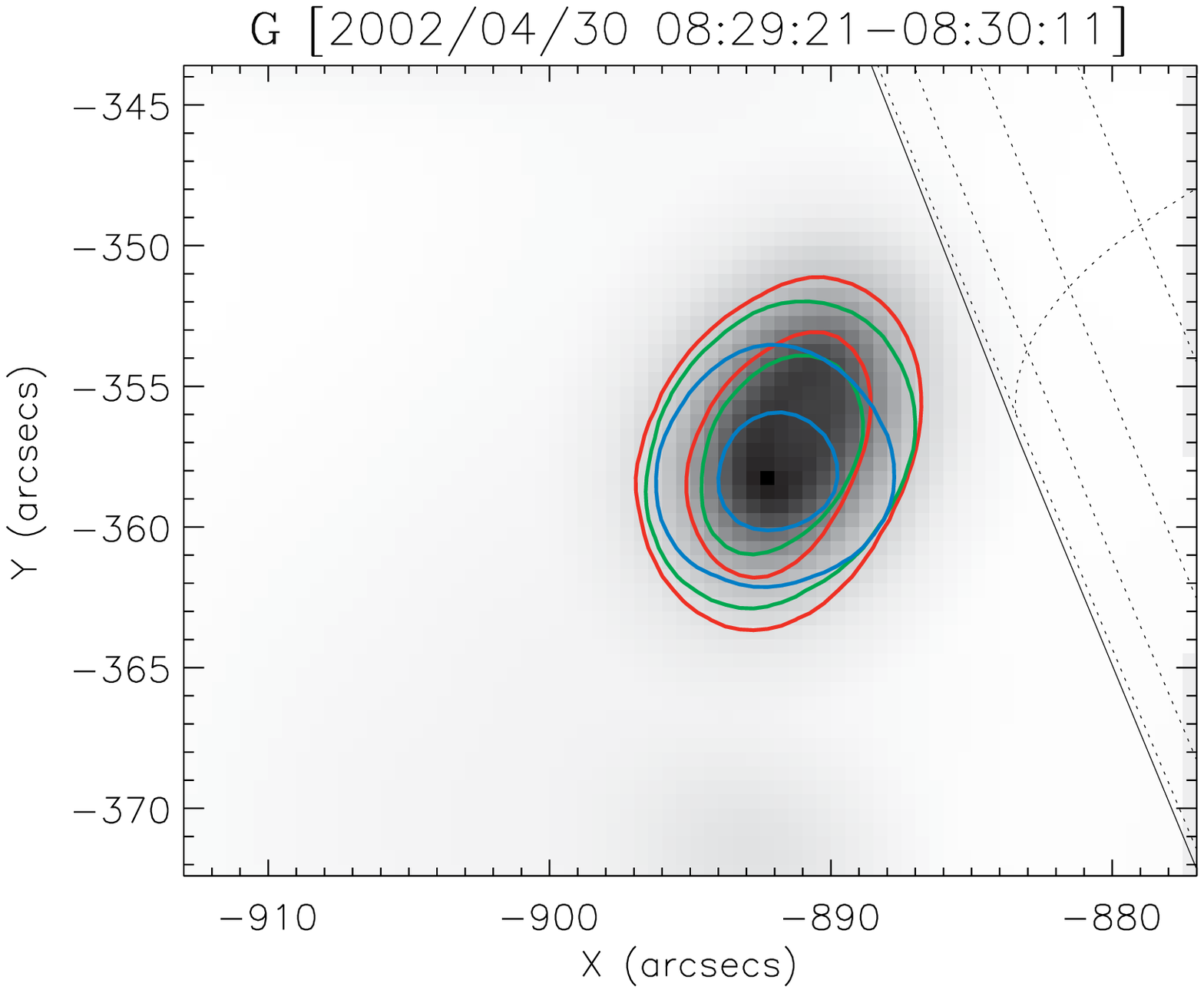}
 \plotone{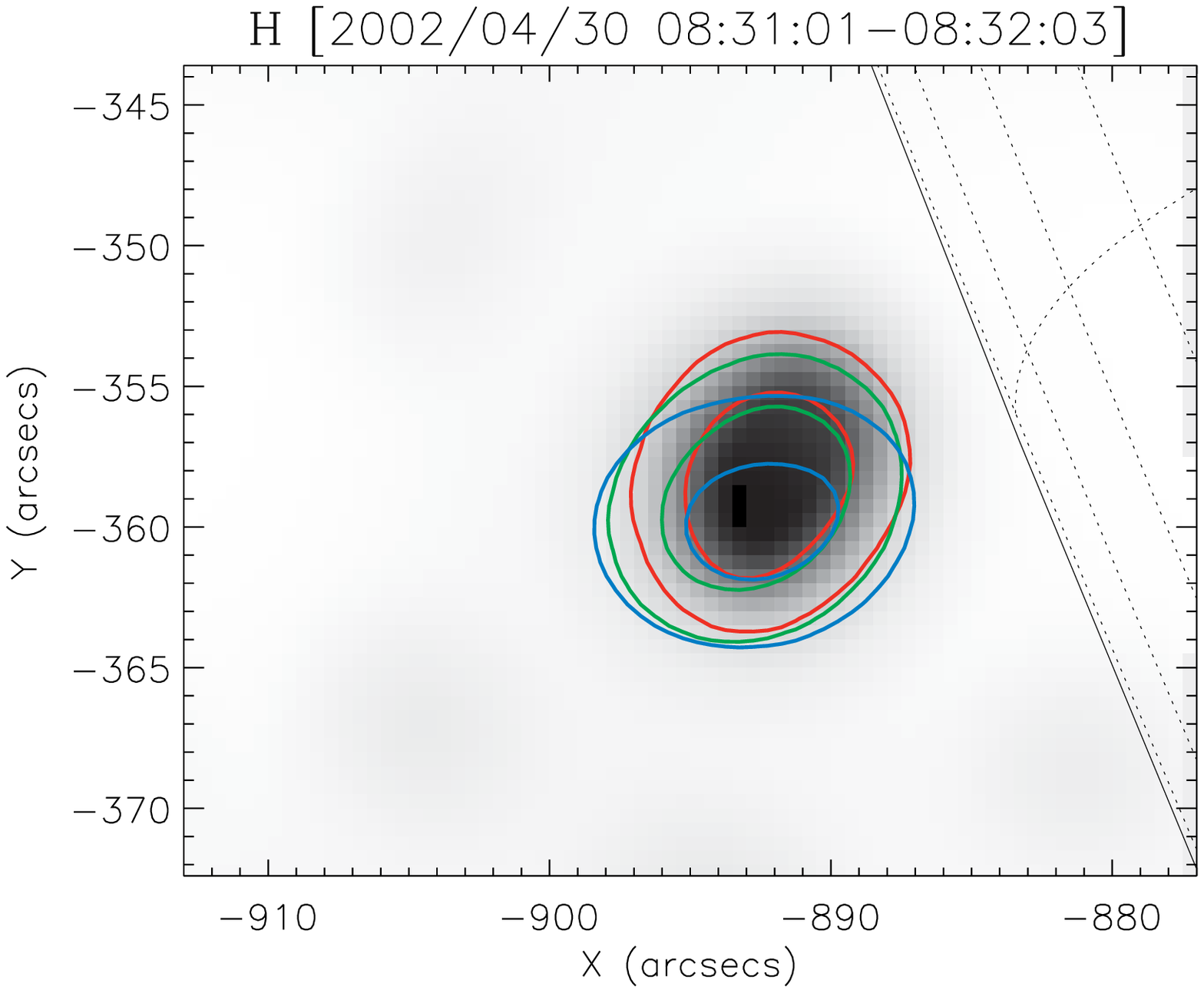}
 \plotone{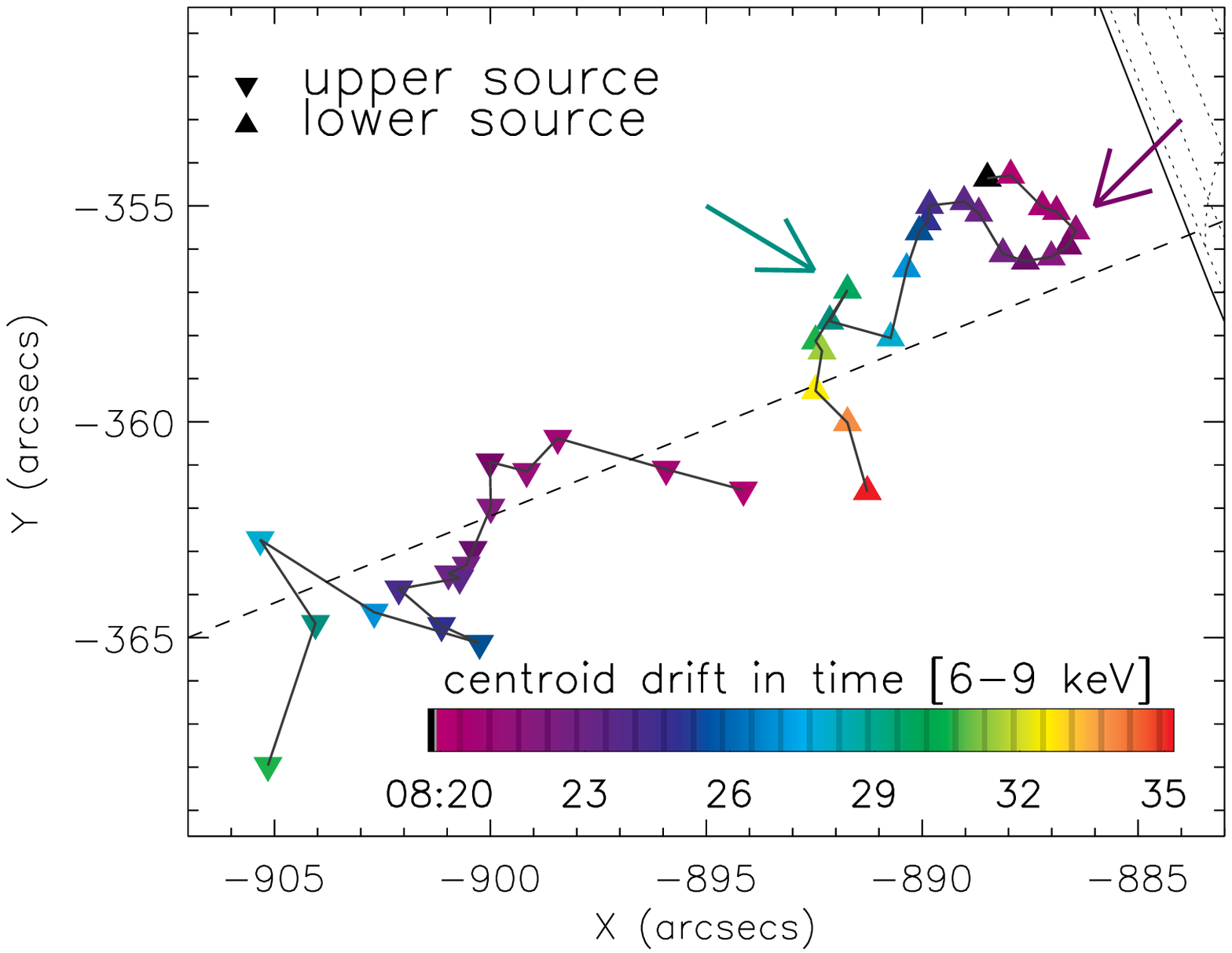}
 \caption[Images of three different energies at selected times]	
 {PIXON images of different energies made with detectors 3--6 and 8 at selected times 
 (i.e., intervals A--H as marked in Figure~\ref{lc.ps}).	
 In each panel, the gray-scale background is at 6--9~keV, while the red, green and blue contours (20\% and 70\% of 
 the peak flux of each image) are at 9--12, 12--16 and 16--25~keV, respectively. The heliographic grid spacing is 2\degree.
 The last panel shows the locations of the centroids of the lower and upper 6--9~keV sources
 at different times indicated by the color bar.	
 The dashed line indicates the radial direction (perpendicular to the limb). 
 The magenta and green arrows point to the centroid locations at the times of the first and second HXR peaks,
 respectively.	
 } \label{chp2LT_mosaic}	
 \end{figure}

Figure~\ref{chp2LT_mosaic} shows the PIXON images taken at 6--9, 9--12, 12--16 and 16--25~keV at eight 
separate times (labeled A--H in Figure~\ref{lc.ps}). 
The morphology evolves following the general trend mentioned above. 
Early (08:19:28-08:20:01~UT, interval A) in the flare, only a single source is visible.
During the next time interval (B), the upper coronal source appears at 6--9~keV, but only a single source is 
evident at higher energies albeit with elongated shapes. In interval C, two distinct coronal sources 
appear in a dumb-bell shape at all the energies.
As time proceeds, both sources move to higher altitudes.	
This morphology is present for about 12 minutes (from 08:20 to 08:32~UT) until the 
declining phase of the second peak when only one source is detected, possibly because 
of the faintness of the upper source and the low count rate.  Note that after 08:29~UT the upper source
is dimmer than 20\% of the maximum of the image and thus does not appear in panels G and H.

The motions of the sources can be seen more clearly from the migration of the centroids.
   To obtain the centroids and fluxes of the sources, we use contours whose levels are equal to 
within 5\% of the minimum between the two sources so that the contours of the two sources are independent.
The last panel in Figure~\ref{chp2LT_mosaic} shows the evolution of the centroid positions of the two sources
at 6--9~keV.  During the first HXR peak (indicated by the {\it magenta} arrow), 
the lower coronal source first shifts to lower altitudes 
and then ascends.  
This is consistent with the decrease of the loop-top height early during the flare observed in several other events
\citep{SuiL2003ApJ...596L.251S, LiuW2004ApJ...611L..53L, SuiL2004ApJ...612..546S}. 
Meanwhile, the upper source generally moves upwards.  Such centroid motions are also present at other energies
as shown in the lower right panel of Figure~\ref{structure}.
The reversal of the lower source altitude seems to happen again, but less obvious, during the second peak
(marked by the {\it green} arrow).

\begin{figure}[thbp]      
 \epsscale{1.1}  
 \plotone{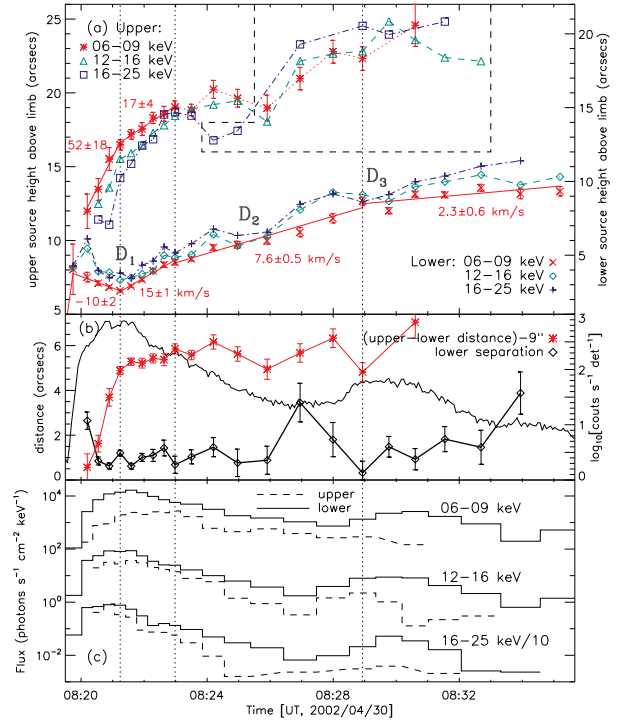}		
 \caption[History of the centroid heights and HXR fluxes of the two coronal sources]
 { ({\it a}) Height (above the limb) of the centroids at different energies for the upper ({\it left scale}) and 
 lower ({\it right scale}) coronal sources. 
 The dotted vertical lines separate the different phases according to the motion of the lower source centroid (see text).
 The red solid lines are linear fits to the data during the corresponding
 time intervals, with the adjacent red numbers indicating the velocities of the altitude gain in units of $\km \ps$.  
 The centroid position of the upper source has large fluctuations and uncertainties during the interval marked
 by the dashed box.  The letters D$_1$, D$_2$, and D$_3$ mark the times when the altitude of the
 lower source decreases. 
    ({\it b}) {\it Left scale}: Distance ({\it red asterisks}) between the centroids of the two coronal 
 sources at 6--9~keV and separation ({\it diamonds})	
 between the centroids of the lower 
 source at 6--9 and 16-25~keV.  The former is shifted downwards by $9 \arcsec$.
     {\it Right scale}: Base-10 logarithm of the spatially integrated light curve (counts~$\ps$~detector$^{-1}$, {\it thin line})
 at 12--25~keV.		
   ({\it c}) Light curves of the upper ({\it dashed}) and lower ({\it solid})
 coronal sources in the energy bands of 6--9, 12--16, and 16--25~keV (divided by 10).
 The same contours (see text) were used to obtain these light curves and the centroid positions in panel {\it a}. 
 } \label{cmpr_hlc.ps}
 \end{figure}

We can examine the same phenomenon more quantitatively by checking the height of the source centroid
as a function of time at different energies. 
   This is shown in Figure~\ref{cmpr_hlc.ps}{\it a}
for the upper ({\it left scale}) and lower ({\it right scale}) coronal sources.
We find that, again, the higher-energy emission comes from lower altitudes for the upper source
and the lower source shows the opposite trend.  The only exception (indicated by the dashed box) 
to this general behavior occurs for the upper source during the late declining phase of first HXR peak 
and during the second and third peaks when there are large uncertainties because of low count rates.  

  At 6--9~keV ({\it red symbols}), the altitude of the lower source first {\it decreases} at a velocity of $10\pm2\km\ps$, 	
while the altitude of the upper source {\it increases} at a velocity of $52 \pm 18 \km \ps$.   
These are indicated by linear fits ({\it red solid line}) during the high flux period.  
This happens during the rising phase (up to 08:21:14~UT) of the first HXR peak 
and is followed by an increase of the altitudes of the 
two sources with comparable velocities ($15 \pm 1 \km \ps$ and $17 \pm 4 \km \ps$ 
for the lower and upper sources, respectively) 		
during the early declining phase (up to 08:22:59~UT). 
  As time proceeds, the two sources generally continue to move	
to higher altitudes. 	
The velocity of the lower source drops to $7.6 \pm 0.5 \km \ps$ until 08:28:56~UT, around the		
maximum of the second HXR peak, and then to $2.3 \pm 0.6 \km \ps$ afterwards.	
The velocity of the upper source also decreases in general, with
some fluctuations most likely due to the large uncertainties mentioned above.
  The relative motion of the two sources can be seen from the temporal variation of the distance 
between their centroids as shown in Figure~\ref{cmpr_hlc.ps}{\it b} ({\it red asterisks}), which undergoes a
fast initial increase and then stays roughly constant at $15\arcsec \pm 1\arcsec$ within the uncertainties.

  At 12--16 ({\it green}) and 16--25~keV ({\it blue}), the centroids have a trend similar to those at 6--9~keV, except
for the lower coronal source during the early rising phase of the first HXR peak.
The initial increase of the height of the ``lower"%
  \footnote{Again, we assign its centroid to the lower source when there is only a single source detected.
  }
source at about 08:20~UT results from the elongation (see the second panel in Figure~\ref{chp2LT_mosaic}) of the single source,
which could be a combination of the lower and upper sources that are not resolved.
The following rapid decrease in height in the next time interval is a consequence of 
the transition from a single-source to a double-source structure as mentioned earlier. 
The upper source, on the other hand, 	
rises more rapidly than at 6--9~keV during the HXR rising and early declining phases.
Its velocity at 16--25~keV, for example, is $32 \pm 3 \km\ps$ during the interval of 08:21:14--08:22:59~UT.	
This energy dependence of the rate of rise is consistent with the general trend of the loop-top
source observed in several other flares \citep{LiuW2004ApJ...611L..53L, SuiL2004ApJ...612..546S}.
We note in passing that, in addition to the first HXR peak (marked with D$_1$ in Figure~\ref{cmpr_hlc.ps}),
the altitudes of the lower source centroids also appear to first decrease and then increase during
two other time periods%
  \footnote{D$_3$ coincides with the second HXR peak 	
  and D$_2$ occurs around 08:25~UT, which is the possible actual start of the second energy release episode
  (see Figure~\ref{lc.ps}), when the upper source also 	
  appears to show a significant decrease in centroid altitude.  Such altitude variations seem to be associated with the possible increases
  of energy release rate indicated by the light curves.
    However, compared with D$_1$, the features at the two later times
  are less definitive given the relatively fewer data points	
  and larger uncertainties of the centroid heights.}
(D$_2$ and D$_3$).  This effect is most pronounced at 12--16 and 16--25~keV.

%



\subsection{Spectral Evolution}
\label{section_spec}

In this section, we examine the relationship between the fluxes and spectra of the two coronal sources.
Figure~\ref{cmpr_hlc.ps}{\it c} shows the photon flux evolution at 6--9, 12--16, and 16--25~keV. 
As evident, the fluxes of the two sources basically follow the same time variation in all three energy 
bands.	
The upper coronal source, however, appears later and disappears earlier, presumably	
due to its faintness and the limited \hsi dynamic range ($\sim$10:1).
It also peaks later at 6--9~keV.

We also conducted imaging spectroscopic analysis for each of the seven time intervals defined
in Figure~\ref{lc.ps}.  The spectra of the two sources separately and the spatially integrated spectra
were fitted with a single-temperature thermal spectrum plus a power-law function.
One important step was to fit the spatially integrated spectra of individual detectors separately 
and then average the results in order to obtain the best-fit parameters and their uncertainties.
Interested readers are referred to Appendix \ref{appendixA} for the technical
details of 	
the spectrum fitting procedures used to obtain the results reported here.
%
%
 \begin{figure}[thbp]      
 \epsscale{0.57}	
 \plotone{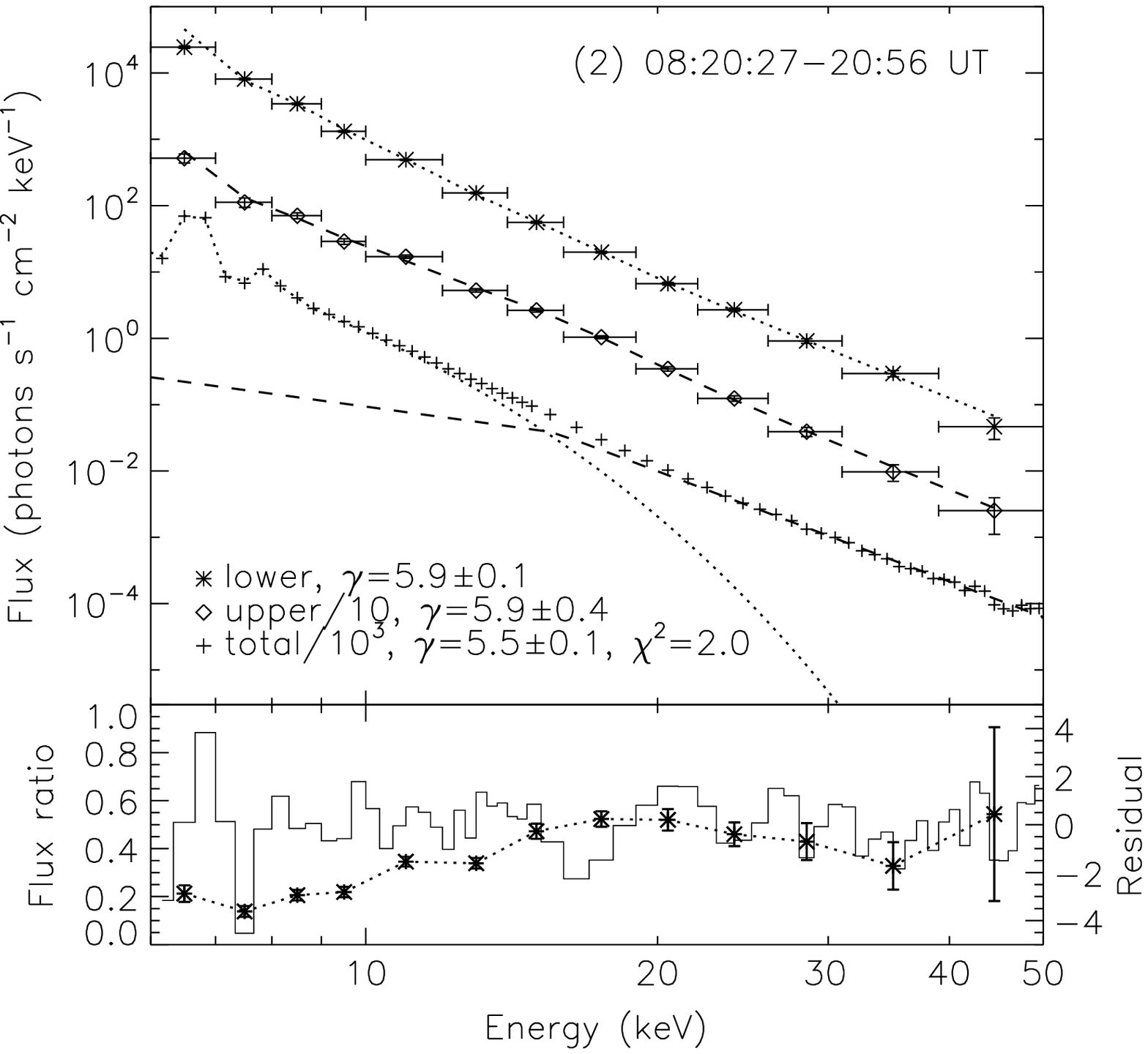}	
 \plotone{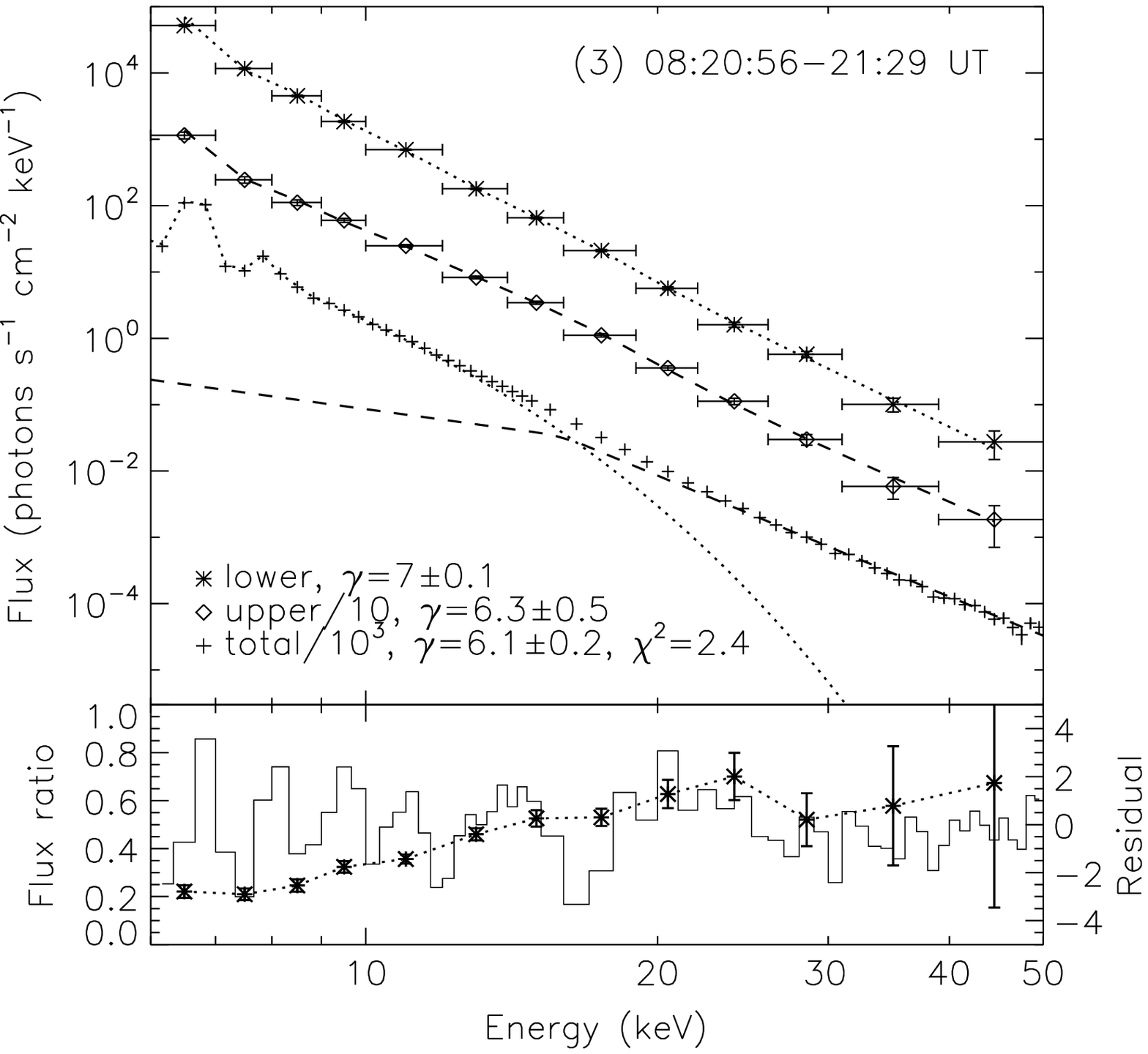}	
 \plotone{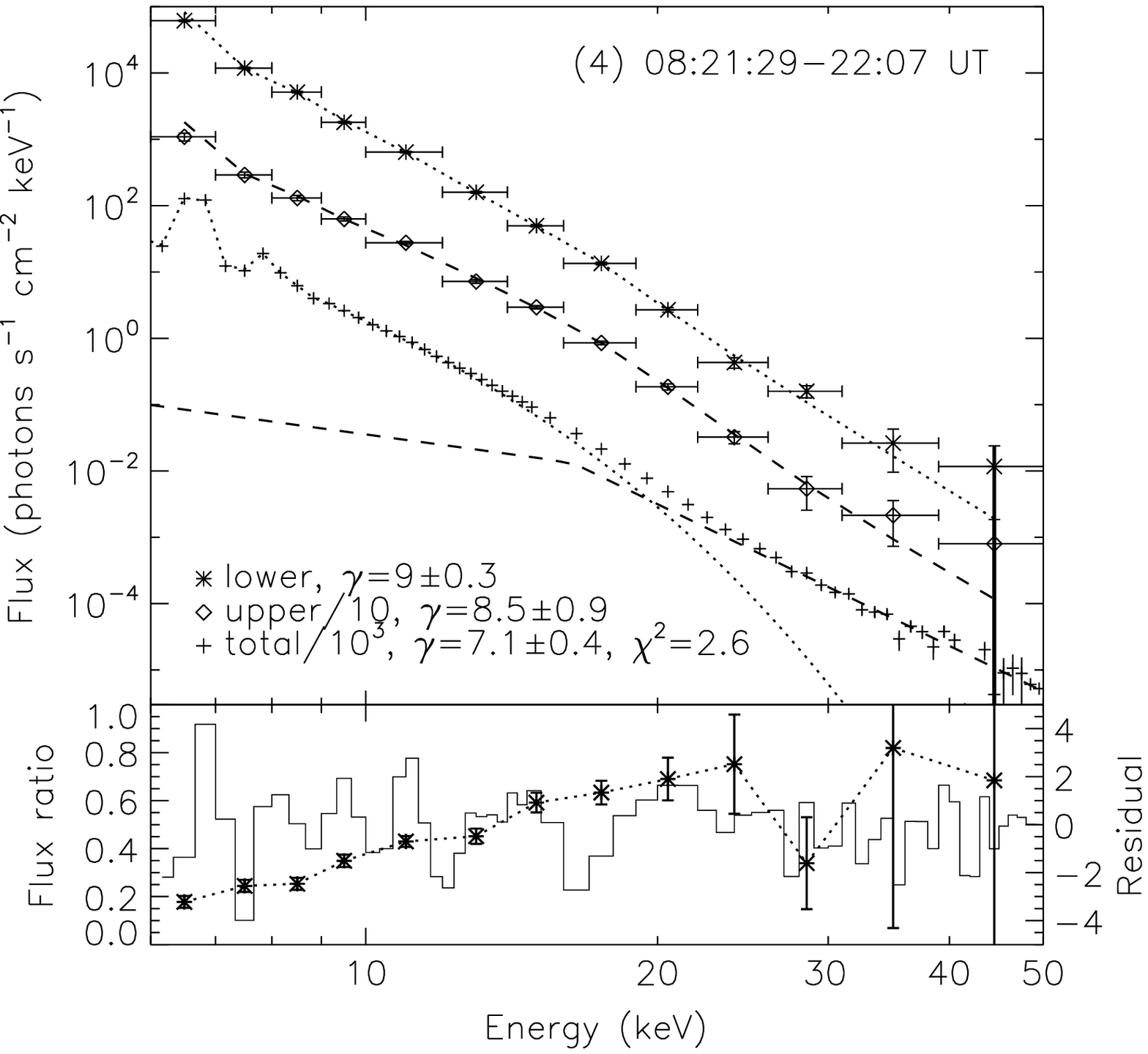}  
 \plotone{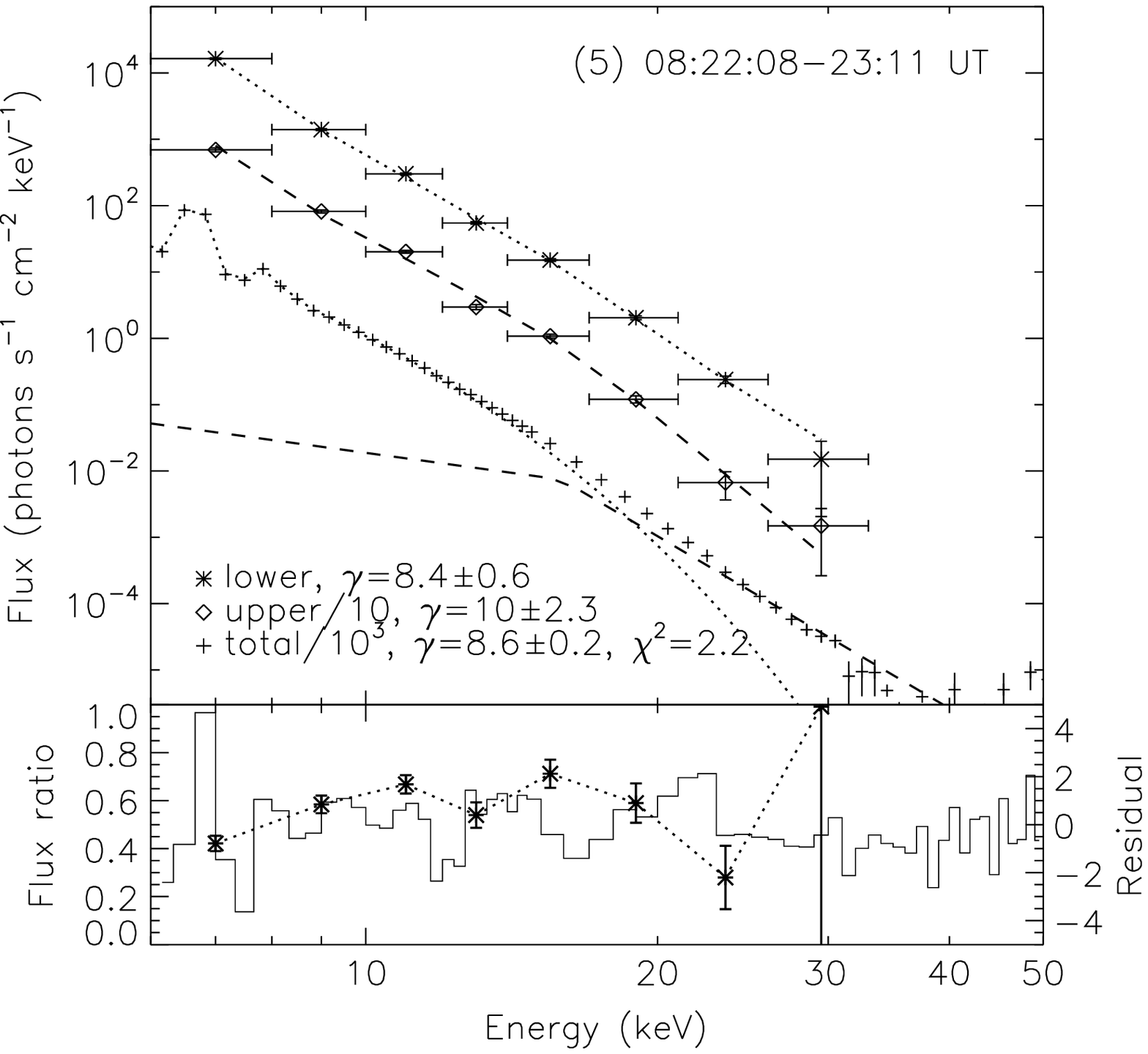}
 \caption[Imaging spectroscopy of the two sources in the 04/30/2002 flare]
 {Spectra of the lower and upper coronal sources and the spatially integrated spectra
 (labeled as ``total") at four times during the major flare peak. 
 The numbers (2, 3, 4, and 5) in the upper-right corners correspond to the numbered 
 time intervals shown in Figure~\ref{lc.ps}.  The upper source's spectra and the total spectra have been 
 shifted downwards by one and three decades, respectively. The horizontal error bars represent the energy bin
 widths and the vertical error bars are the statistical uncertainties of the spectra. 
 The best fit to the data with a thermal plus power-law model is shown as the {\it dotted} ({\it dashed}) 
 line for the lower (upper) source.  The thermal ({\it dotted}) and power-law ({\it dashed}) components 
 of the best fit to the total spectra are also shown. The legend indicates the corresponding power-law 
 indexes ($\gamma$) for each spectrum ($\gamma=2$ below the low cutoff energy).  
   The lower portion of each panel shows the ratio of the upper to lower fluxes ({\it asterisk} symbols, {\it left scale}),
 and the residuals ({\it solid} lines, {\it right scale}) of the fit to the spatially
 integrated spectra, normalized to the $1\sigma$ uncertainty of the measured flux at each energy.
 } \label{chp2LT_imgspec}
 \end{figure} 	

%

\begin{figure}[thbp]      
 \epsscale{0.9}	
 \plotone{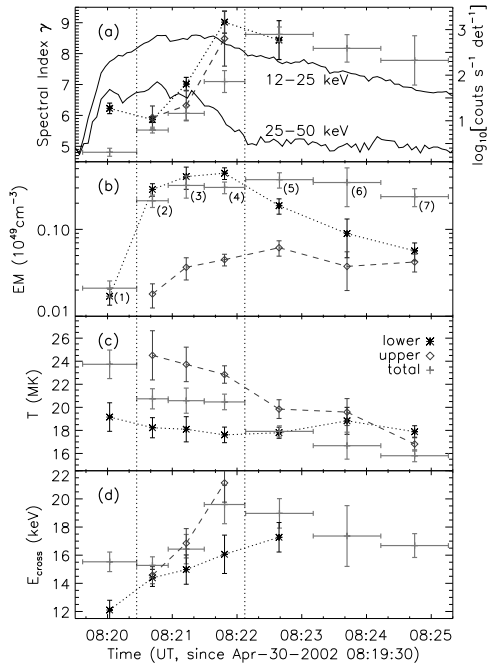}	
 \caption[]
 {Evolution of various spectroscopic quantities of the lower ({\it asterisks}) and upper ({\it diamonds}) 
 coronal sources and the spatially integrated emission ({\it pluses}, labeled ``total"). The horizontal 
 error bars represent the widths of the time intervals of integration as labeled (1--7) in panel {\it b} 
 (also in Figure~\ref{lc.ps}).  The two vertical dotted lines mark the boundaries of the time range when
 both coronal sources are best imaged.  
 This spans the late 
 impulsive phase (see the 25--50~keV light curve). Before and after this time range the imaging spectroscopy 
 has relatively large uncertainties (see text).
   ({\it a}) Spectral indexes ({\it symbols}, {\it left scale}) of the power-law components of the model fits, together with
 the 12--25 and 25--50~keV light curves ({\it solid lines}, {\it right scale}).   
   ({\it b}) and  ({\it c}) Emission measures (in $10^{49} \pcmc$) and temperatures (in $10^6$~K) of 
 the thermal components of the model fits.
   ({\it d}) The cross-over energy, $E_\unit{cross}$, at which the thermal and power-law components are equal.	
 Note that the values here are the upper limits of $E_\unit{cross}$.  This is because we assumed 
 a $\gamma=2$ index for the photon spectrum below the low-energy cutoff, but the power-law component may 
 extend to low energies with a steeper index thus lowering the values of $E_\unit{cross}$.
 } \label{fig_index}
 \end{figure}

A sample of the resulting spectra of four intervals is shown in Figure~\ref{chp2LT_imgspec}. 
Fits to the spatially integrated spectra indicate that the low-energy emission is dominated 
by the thermal components, while the nonthermal power-law components dominate at high energies.  The two components
cross each other at an energy that we call $E_\unit{cross}$. 
The spectra of the two coronal sources measured separately have similar slopes.	
In general, the ratio of the two spectra (upper source/lower source) is smaller than unity and 
gradually increases with energy below around $E_\unit{cross}$.  This trend can also be appreciated 
by noting the increasing relative brightness of the upper source when energy increases 	
as shown in Figure~\ref{hsi_multi.ps}.
This energy-dependent variation of the flux ratio means that the thermal emissions of the two sources are 
somewhat different not only in emission measure (EM) but also in temperature, because 
different EMs alone would only affect the normalizations and produce a 	
flux ratio that is independent of energy.
We also note that above $E_\unit{cross}$, the ratio stays 	
constant within the larger uncertainties.
This means that the nonthermal spectra of the two sources have similar power-law indexes 
(see Figure~\ref{fig_index}{\it a}).

The reduced $\chi^2$ values of the spatially integrated spectra are somewhat large ($\gtrsim$2) 
partly because we set the systematic uncertainties to be zero as opposed to the default 2\%.
Another reason was that we averaged the photon fluxes and best-fit parameters over different detectors 
that have slightly different characteristics.  Thus, the averaged model may not necessarily be 
the best fit to the averaged data (see \S\ref{appendix_spec}, item 9), 
although the $\chi^2$ values of the fits to the individual detectors are usually close to unity.
  The normalized residuals exhibit some systematic (non-random) variations, as shown in the bottom portion of 
each panel of Figure~\ref{chp2LT_imgspec}.  This suggests that simple spectrum form adopted here may not
represent all the details of the data.  
However, since we are mainly concerned with the similarities and differences
between the spectra of the two coronal sources, such systematic variations would affect both spectra
the same way and thus will not alter our major conclusions.
  More sophisticated techniques, such as the regularization method \citep{KontarE2004SoPh.regularize}, 
can be used to obtain better fits to the data, but they are beyond the scope of this paper.

%
We now examine the temporal evolution of various spectral characteristics as shown in Figure~\ref{fig_index}.
Let us focus on the late impulsive phase outlined by the two vertical dotted lines%
  \footnote{Beyond the time interval between the two vertical lines in Figure~\ref{fig_index}, i.e., 
  during the early impulsive phase (before 08:20:27~UT) and the decay phase (after 08:22:08~UT), 
  interpretation of the spectral fitting needs to be taken with caution because of the large 
  uncertainties due to low count rates and thus relatively poor statistics.  
  Specifically, during certain intervals, reliable power-law components from fits to the spatially resolved
  spectra could not be obtained and thus the corresponding values of the spectral index ($\gamma$)
  and thermal-nonthermal cross-over energy ($E_\unit{cross}$) are not shown in Figure~\ref{fig_index}.
  In addition, the averages of the best-fit parameters of the two sources 
  differ significantly from the corresponding values of the spatially integrated spectrum.
  This is unexpected and may indicate that there existed an extended source with low-surface brightness and/or
  that the fits to the imaged spectra at these times are not reliable.
  Nevertheless, we show the fitting results here for completeness. 
  }.
Again, we find that the power-law indexes (Figure~\ref{fig_index}{\it a}) of the two coronal sources 
are very close, with a difference of $\Delta\gamma \leq 0.7$. 		
The two spectra undergo continuous softening during this stage, 
and the spatially integrated spectrum follows the same general trend.

Figures~\ref{fig_index}{\it b} and \ref{fig_index}{\it c} show that 
the thermal emissions of the two sources are quite different as noted above.
The lower coronal source has a larger emission measure but lower temperature than the upper source.  
  As time proceeds, both sources undergo a temperature decrease and emission measure increase. 
This must be the result of the interplay of continuous heating, cooling by conduction and radiation,
and heat exchange  
between regions of different temperatures within the emission source.  
Note that the temperature and emission measure of the spatially integrated spectrum, as expected, lie 
between those of the two sources.	

We can further estimate the densities of the two sources using their EMs and approximate volumes.
Assuming that the sources are spheres and using the $6 \farcs 3$ and $5 \farcs 2$ FWHM source sizes
obtained from the visibility forward fitting images as the diameters, we obtained the volumes, $V$.
We then estimated the lower limits of the densities ($n=\sqrt{\unit{EM}/[V f]}$, 
assuming a filing factor $f$ of unity) of the lower and upper sources at 08:20:27-08:20:56~UT 
to be $2.4 \times 10^{11}$ and $8.0 \times 10^{10}\pcmc$, respectively.

  Figures~\ref{fig_index}{\it d} shows the history of the cross-over energy $E_\unit{cross}$.
In general, the lower source has a lower $E_\unit{cross}$ because of its lower temperature.
The $E_\unit{cross}$ values of both sources increase with time because the thermal emission becomes
increasingly dominant, as seen in many other flares. 	
Physical interpretation of these observations is presented in the next section.


\section{Interpretation and Discussion}		
\label{section_interpret}


\subsection{Energy Dependence of Source Structure}	
\label{interp_morphovsE}

The energy-dependent source morphology presented in \S \ref{chp2LT_morphovsE} 
(see Figure~\ref{structure}, {\it lower left}) is similar to that reported by 
\citet{SuiL2003ApJ...596L.251S} and \citet{SuiL2004ApJ...612..546S}
and interpreted as magnetic reconnection taking place between the two coronal sources. In their interpretation,
plasma with a higher temperature is located closer%
  \footnote{\citet{SuiL2003ApJ...596L.251S} also suggested a possible transition at about 17~keV from the 
  thermal flare loops to the Masuda-type above-the-loop HXR source \citep{Masuda1994Nature}, 
  on the basis of the sudden displacement of the loop-top source position in the 2002 April 15 flare.}
to the reconnection site than plasma with a lower temperature.
This can result in higher-energy emission coming from a region closer to the reconnection site
while lower-energy emission comes from a region further away, 
provided that the emission is solely produced by {\it thermal} 
emission (free-free and free-bound) and the lower-temperature plasma has a higher emission measure.

\begin{figure}[thbp]      
 \epsscale{1.2}	
 \plotone{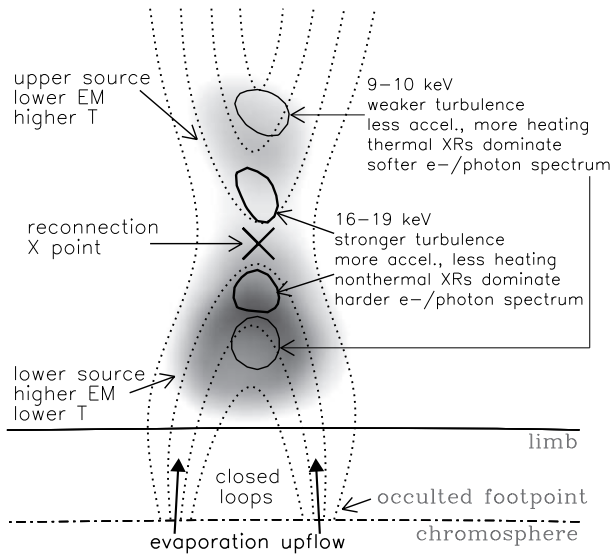}
 \caption[Cartoon]
 {Schematic of the physical scenario (see text) superimposed on the \hsi observations, as a manifestation
 of the stochastic acceleration model illustrated in Figure \ref{fig_model}. 
 The 14--16~keV PIXON image at 08:20:27--08:20:56~UT is the {\it gray background}, 
 overlaid with the simultaneous 9--10 ({\it thin}) and 16--19~keV ({\it thick}) contours. 
 These are the same images shown in the lower left panel of Figure~\ref{structure}, with their orientation 
 rotated for demonstration purposes.  The hand-drawn dotted curves represent a possible magnetic field configuration.
 } \label{fig_cartoon}
 \end{figure}

Our interpretation is somewhat different, particularly for this flare. 
  Regardless of the emission nature (thermal or nonthermal) of the HXRs, 
the energy-dependent source structure here simply means {\it harder} (flatter) 
photon spectra {\it closer} to the reconnection site, which can give rise to 
a higher weighting there at high energies for the centroid calculations.
A larger spatial gradient of the spectral hardness would lead to a larger separation of
the emission centroids at two given photon energies, and a zero gradient (uniform spectrum) 
means no separation.
  As we have seen in \S\ref{section_spec}, both coronal sources have substantial power-law (presumably nonthermal) tails 
(Figure~\ref{chp2LT_imgspec}), which makes a purely thermal interpretation improbable.  
  In the framework of the stochastic acceleration model \citep{HamiltonR1992ApJ...398..350H, MillerJ1996ApJ...461..445M}, 
one expects both heating of plasma and acceleration of particles into a nonthermal tail to take place.
As shown in \citet{PetrosianV2004ApJ...610..550P}, higher levels of turbulence tend to produce harder
electron spectra or more acceleration and less heating. 
  One expects a higher turbulence level near the X-point of the reconnection site than further away. 
Consequently, there will be more acceleration and thus stronger nonthermal emission near the center, 
but more heating and thus stronger thermal emission further away from the X-point.  
In other words, the electron spectra and thus the observed photon spectra will be harder 
closer to the reconnection site.  This physical picture is sketched in Figure~\ref{fig_cartoon}.

The observations here support the above scenario.  As shown in Figure~\ref{chp2LT_imgspec},		
below the critical energy $E_\unit{cross}$ (say, $\sim$15~keV for 08:20:27--08:20:56~UT), 
the emission is dominated by the thermal component, and the two sources are further apart at 
lower energies (see Figure~\ref{structure}, {\it lower} panels).  This translates to the outer region away from the 
center of the reconnection site being mainly thermal emission at low energies.
Above $E_\unit{cross}$, on the other hand, the power-law component dominates. 
The two sources being closer together at higher energies%
  \footnote{At even higher energies ($\gtrsim$25~keV), the distance between the two sources 
  seems to increase, but with larger uncertainties (Figure~\ref{structure}, {\it lower right}). 
  This transition, if real, may suggest that transport effects become important.
  This is because higher-energy electrons require greater column depths to stop them, 
  and thus they tend to produce nonthermal bremsstrahlung emission at larger distances from where they are accelerated.}
thus means that the region near the center is dominated by nonthermal emission%
  \footnote{This argument is equivalent to the approach of obtaining detailed spectroscopy of 
  multiple regions as small as $4\arcsec$, but this was not attempted here.
  } 
(see Figure~\ref{fig_cartoon}).
	%

We note that the small centroid separation of $4\farcs 6 \pm 0\farcs 3$ (Figure~\ref{structure}, {\it lower left})
identifies a region within which the center of reconnection activity is located.  To our knowledge, this is
the smallest ($3.3\pm0.2$~Mm) feature of the reconnection region yet resolved by X-ray observations on the Sun. 

\subsection{Temporal Evolution of Source Structure}	
\label{section_time_02}

Figure~\ref{cmpr_hlc.ps}{\it b} shows the separation ({\it black diamonds}) between the centroids 
of the lower coronal source at 6--9 and 16--25~keV, together with the 12--25~keV light curve.
These two curves seem to be anti-correlated such that this separation becomes smaller
when the HXR flux is larger.  This is consistent with that reported by
\citet[][Figure~1]{LiuW2004ApJ...611L..53L} in a much brighter (X3.9) flare where
this effect was more pronounced.  This trend 	
was also present in two of the three homologous flares reported by \cite{SuiL2004ApJ...612..546S}.

  In our earlier publication \citep{LiuW2004ApJ...611L..53L}, we suggested that
the anti-correlation indicates a smaller (more homogeneous) spatial gradient of 
turbulence density or particle acceleration rate around the peak of the impulsive phase, 
owing to the presence of a higher turbulence level.	
Here, we further note that such a spatial distribution of acceleration rate can
result from the interplay of various physical processes (with different spatial distributions and time scales)
that contribute to energy release, dissipation, and redistribution.  Processes that can carry energy away from the 
acceleration region include damping of turbulence (waves), escape of accelerated particles, 
thermal conduction, and radiative loss.  Detailed modeling is required to offer a self-contained
physical explanation for the observational feature presented here.

\subsection{Spectral Characteristics}
\label{interp_spec}

The temporal {\it correlation} of the light curves (Figure~\ref{cmpr_hlc.ps}{\it c}) 
and the {\it similar} power-law spectral components (Figures~\ref{chp2LT_imgspec} and \ref{fig_index}{\it a})
of the two coronal sources, when taken together, suggest that these HXR emissions are produced by the 	
nonthermal electrons that are accelerated by the {\it same} mechanism (presumably stochastic acceleration 
by 	
similar turbulence) following the reconnection process.
Such a correlation provides more direct evidence and a more complete picture for the interpretation 
outlined above in \S\ref{interp_morphovsE}.

As we noted in \S \ref{section_spec}, the coronal sources have quite different {\it thermal} emissions,
with the lower source having a higher EM but lower temperature.
There are several possible reasons why this can happen: 
  (1) The lower source resides at a lower altitude where the local density may be slightly 
higher in the gravitationally stratified atmosphere.   
  The difference between the heights of the two sources is on the order of 10~Mm, which
  is a fraction of the coronal density scale height ($\gtrsim$60~Mm, the quiet Sun value).
  Thus, the density difference due to height variation is no more than about 15\%.
This is not sufficient to account for the large difference in density
between the two sources noted earlier.	
  (2) As shown in Figure~\ref{fig_cartoon}, the two coronal sources lie below and above the X point 
of the reconnection region.  It is most likely that the lower source is located at the top 
of the flaring loop that is magnetically connected to the chromosphere.
This allows the chromosphere to supply dense material to the lower source 
along the magnetic field lines during chromospheric evaporation.  The evaporated plasma, although heated,
is still relatively cooler than the hot plasma near the reconnection site in the corona.
  (3) In addition, thermal conduction and plasma convection can readily carry heat away from the lower source
down the magnetic loop to the cool chromosphere.
All the three reasons contribute to the higher EM and lower temperature of the lower coronal source.
In contrast, the upper source may be magnetically disconnected from or more remotely connected to
the solar surface. 
The lower density material of the upper source can thus be heated to a higher temperature 
due to the lack of a direct supply of cool material and the reduced thermal conduction to 	
the chromosphere.  

We have also noted that the {\it nonthermal} components of the two coronal sources have similar spectral
indexes, 	
but the upper source is weaker.
The spectral indexes could not always be determined for both sources seen in other similar events
\citep{SuiL2003ApJ...596L.251S, SuiL2004ApJ...612..546S, VeronigA2006A&A...446..675V, LiYPGanWQ2007AdSpR},
but the upper source was always the weaker of the two.
Here we discuss the possibilities that can lead to the weaker nonthermal radiation
of the upper source in particular and its low surface brightness in general.
In the framework of the stochastic acceleration model, all the processes involved in producing the 
observed emission -- the rate of generation of turbulence, the spectrum of turbulence, 
the rate of acceleration and emission -- depend on the temperature, density, 
and the magnetic field strength and geometry.
As we mentioned above, the temperature, density, and field geometry of the two coronal sources are different.
The magnetic field strength most likely decreases with height.	
Consequently, we expect different HXR intensities from the two sources.
   For example, the lower plasma density in the upper source will result
in lower surface brightness for both thermal and nonthermal bremsstrahlung emission.
   Magnetic topology can have similar effects. The electrons
responsible for the upper source are likely to be on open field lines 
or on field lines 
that connect back to the chromosphere more remotely \citep[e.g.,][]{LiuC2006ApJ}
and thus produce their X-ray
emission in a more spatially diffuse region.  In contrast, for the lower
source, the electrons are confined in the closed loop. 	
In addition, as noted above, chromospheric evaporation can
further increase the density in the loop, enhancing the density
effect mentioned here. These factors, again, lead to lower surface brightness for the upper source.
  Finally, 	
the rate of acceleration or heating depends primarily
on the strength of the magnetic field \citep{PetrosianV2004ApJ...610..550P},
so that the relatively weaker magnetic field of the upper source may  	
result in slower acceleration and thus weaker nonthermal emission.  A large sample of this
type of flares is required to confirm or reject this explanation.

We should emphasize that since the 	
radiating electrons in both sources are the {\it direct} product of 
the same acceleration mechanism, they share common signatures. 
 This would explain the spectral similarity of the nonthermal
emissions of the two coronal sources.  The thermal X-ray emitting plasma, however, in addition
to direct heating by turbulence, involves many other {\it indirect} or {\it secondary} processes, 
such as cooling by thermal conduction and hydrodynamic effects (e.g., evaporation in the closed loop).
Therefore, the two thermal sources exhibit relatively large differences 
in their temperatures and emission measures.


\section{Concluding Remarks}		
\label{chp2LT_summary}


We have performed imaging and spectral analysis of the \hsi observations of the M1.4 flare that occurred 
on April 30, 2002.  Two correlated coronal HXR sources appeared at different altitudes
during the impulsive and early decay phases of the flare.  
The long duration ($\sim$12 minutes) of the sources allows for detailed analysis
and the results  	
support that	
magnetic reconnection and particle acceleration were taking place between the two sources. 
Our conclusions are as follows.

\begin{enumerate}


 \item Both coronal sources exhibit energy-dependent morphology. Higher-energy emission comes from higher
altitudes for the lower source, while the opposite is true for the upper source 
(Figures~\ref{hsi_multi.ps} and \ref{structure}). 
This suggests that the center of magnetic reconnection is located 
within the small region between the sources.

 \item The energy-dependent source structure (Figure~\ref{structure}), combined with spectrum analysis (Figure~\ref{chp2LT_imgspec}), 
implies that the inner region near the reconnection site is energetically dominated by nonthermal emission,
while 	
the outer region is dominated by thermal emission.  
This observation, in the framework of the stochastic 
acceleration model developed by \citet{HamiltonR1992ApJ...398..350H} and \citet{PetrosianV2004ApJ...610..550P},
supports the scenario  (Figure~\ref{fig_cartoon}) that a higher turbulence level and thus more acceleration 
and less heating are located closer to the reconnection site.

 \item The light curves (Figure~\ref{cmpr_hlc.ps}{\it c}) 	
and the shapes of the nonthermal spectra (Figures~\ref{chp2LT_imgspec} and \ref{fig_index}{\it a}) 
of the two X-ray sources obtained from
imaging spectroscopy are similar.	
This suggests that   
intimately related populations of electrons, presumably heated and accelerated
by the same mechanism following energy release in the same reconnection region,	 
are responsible for producing both X-ray sources.

 \item The thermal emission indicates that the lower coronal source has a larger emission measure but lower
temperature than the upper source (Figures~\ref{fig_index}{\it b} and \ref{fig_index}{\it c}).  This is ascribed to 
the expected different magnetic connectivities of the two sources with the solar surface  
and the associated different plasma densities.	

 \item During the rising phase of the main HXR peak,
the lower source (at 6--9~keV) moves {\it downwards} for nearly two minutes at a velocity of $10\pm2\km\ps$,
while the corresponding upper source moves {\it upwards} at $52 \pm 18 \km \ps$ (Figure~\ref{cmpr_hlc.ps}{\it a}).
During the early HXR declining phase, the two sources move upwards at comparable velocities
($15\pm 1\km \ps$ vs. $17 \pm 4 \km \ps$) for another two minutes. 
Afterwards, both sources generally continue to move upwards with gradually
decreasing velocities throughout the course of the flare, with some marginally significant fluctuations.

 \item For 	
the lower source, the separation between the centroids of the emission at different energies 
seems to be anti-correlated with the HXR light curve (Figure~\ref{cmpr_hlc.ps}{\it b}), which is
consistent with our earlier finding \citep{LiuW2004ApJ...611L..53L}.
In the stochastic acceleration model, such a feature suggests that a stronger 
turbulence level (thus a larger acceleration or heating rate and a higher HXR flux) is associated 
with a smaller spatial gradient (i.e., more homogeneous)		
of the turbulence distribution or of the electron spectral hardness.

\end{enumerate}

All the above conclusions fit the picture of magnetic reconnection taking place between the two
sources as illustrated in Figure~\ref{fig_cartoon}.  This is another, yet stronger, 
case of a double-coronal-source morphology observed in X-rays, in addition to the five other events 
reported by \citet{SuiL2003ApJ...596L.251S}, \citet{SuiL2004ApJ...612..546S}, \citet{VeronigA2006A&A...446..675V},
and \citet{LiYPGanWQ2007AdSpR}.  	


The general variation with height of the coronal emission raises some
interesting questions and provides clues to the energy release
and acceleration processes.

The fact that there are two sources rather than one elongated continuous source
suggests that 	
energy release takes place primarily
away from the ``X" point of magnetic reconnection. This can be explained by the following scenario. One
may envision that the reconnection gives rise to an electric field which results
in runaway beams of particles. This is an unstable situation and will lead 		
to the generation of plasma waves or turbulence, which can then heat and accelerate
particles some distance away from the ``X" point.

In addition, the energy-dependent structure of each source (i.e., higher energy emission being
closer to the ``X" point) that extends over a region of $\lesssim10\arcsec$ suggests that
energy release and some particle acceleration occurs	
in this region.  	
This also indicates
that the turbulence level or acceleration rate decreases with distance from the 
``X" point, which results in softer electron spectra further away from that point. 
    In other words, this observation 
suggests that the usually observed loop-top source is part of the acceleration region that
resides in the loop and has some spatial extent,		
which is consistent with the recent study
reported by \citet{XuY2007ApJ.fwdfit}. 		
(In their cases, the second coronal source at even higher altitudes above the reconnection site were not
detected presumably because of the low total intensity and/or surface brightness.)

Our conclusions do not support the idea that particles are accelerated outside the HXR source	
before being injected into the loop.
Moreover, the observations here are contrary to the predictions of the collisional thick-target model
\citep[e.g.,][]{BrownJ1971SoPh, PetrosianV1973ApJ...186..291P}, which has been generally
accepted for the footpoint emission and was recently invoked by
\citet{VeronigBrown2004ApJ...603L.117V} to explain the bulk coronal HXRs in two 
flares described by \citet{SuiL2004ApJ...612..546S}.
In such a model, one expects higher-energy emission to come from larger distances from the
acceleration site \citep[e.g., see][for HXRs from the legs and footpoints of a flare loop]
{LiuW2006ApJ...649.1124L} due to the transport effects mentioned in \S 3.1.
The electron spectrum becomes progressively harder with distance 
(because low-energy electrons lose energy faster). This disagrees with the observations of the flare
presented here and of the two flares reported by \citet{SuiL2004ApJ...612..546S}.

    We note in passing that there is a common belief that the ``Masuda" type of ``above-the-loop" 
sources \citep{Masuda1994Nature} constitutes a special class of HXR emission. 
We should point out that the ``Masuda" source is most likely an extreme case of
the lower coronal source observed here and of the commonly observed loop-top sources 
that exhibit harder spectra higher up in the corona 
\citep[e.g.,][]{SuiL2003ApJ...596L.251S, LiuW2004ApJ...611L..53L, SuiL2004ApJ...612..546S}. 
 We also emphasize that some type of trapping is required to confine high-energy electrons 
in the corona while allowing some electrons to escape to the chromosphere (see Figure \ref{fig_model}).
Coulomb collision in a high-density corona
cannot explain simultaneous high-energy 	
coronal and footpoint emission at energies as high as 33--54~keV in the Masuda case. 
The stochastic acceleration model, on the other hand, provides   
the required trapping by turbulence 
that can scatter particles and accelerate them at the same time
\citep{PetrosianV2004ApJ...610..550P, JiangY2006ApJ...638.1140J}.


  Finally, besides the stochastic acceleration model, other commonly cited mechanisms, such as acceleration by
shocks \citep[e.g.,][]{TsunetaS1998ApJ...495L..67T}  %
 and/or DC electric fields \citep[e.g.,][]{HolmanG1985ApJ...293..584H, BenkaS1994ApJ...435..469B}, 
may or may not be able to explain the energy-dependent source structure
presented here. 
%
A rigorous theoretical investigation of these models is required to evaluate their viability.



\acknowledgments
This work was supported by NASA grants NAG5-12111, NAG5 11918-1, and NSF grant ATM-0312344 at
Stanford University. WL was also supported in part by an appointment to the NASA Postdoctoral Program at
Goddard Space Flight Center, administered by Oak Ridge Associated Universities through a contract
with NASA. 
We are grateful to Gordon Holman and the referee for critical comments. We also thank Siming Liu, Hugh
Hudson, Tongjiang Wang, Astrid Veronig, and Saku Tsuneta for fruitful discussions, and Kim Tolbert, 
Richard Schwartz, and many other \hsi team members for their invaluable technical support.
WL is particularly indebted to Dr. Thomas R. Metcalf, who tragically passed away
recently, for his help with the PIXON imaging technique.



\appendix

\section{Appendix: \hsi Spectral Analysis}
\label{appendixA}

We document in this section the specific procedures adopted
to obtain the spatially integrated spectra throughout the
flare and the spatially resolved spectra of the two
individual coronal sources during the first HXR peak.
These procedures are refinements to
the standard \hsi image reconstruction \citep{HurfordG2002SoPh..210...61H}	
and spectral fitting \citep{SmithD2002SoPh..210...33S}	
techniques that are implemented in the 	
Interactive Data Language (IDL) routines	
available in the SolarSoftWare \citep[SSW,][]{FreelandSL&HandyBH1998SoPh..182..497F}. 	
Specific analysis routines
are described by \citep{SchwartzRA2002SoPh..210..165S}	
and in various documents on the \hsi web site at
\hbox{http://hesperia.gsfc.nasa.gov/rhessidatacenter/.}
The procedures described here can be readily adopted for 
general \hsi spectral analysis tasks.

\subsection{Spatially Integrated Spectra}
\label{appendix_spec}

For the spatially integrated spectra, we used the standard
forward fitting method implemented in the object-oriented
routine called Object Spectral Executive (OSPEX) and described in
\citet{BrownJ2006ApJ...643..523B}. 
OSPEX uses an assumed parametric form of the photon spectrum
and finds parameter values that provide the best fit in a
$\chi^2$	
sense to the measured count-rate spectrum in
each time interval. 

In analyzing the \hsi spatially integrated count-rate
spectra, we took advantage of the fact that \hsi makes
nine statistically independent measurements of the same
incident photon spectrum with its nine nominally identical
detectors. By analyzing the data from each detector
separately, up to nine values can be obtained for each
spectral parameter. The scatter of these values about the
mean then gives a more realistic measure of the uncertainty
than can be obtained from the best fit to the spectrum
summed over all detectors. In addition, treating each
detector separately allows us to use the 1/3~keV wide	
``native" energy bins of the on-board pulse-height analyzers
for each detector. This avoids the energy smearing inherent
in averaging together counts from different detectors that
have different energy bin edges and sensitivities. We
limited the total number of energy bins by using the
1/3~keV native bins only where they are needed, i.e., between 3 and 15~keV.
This provides the best possible
energy resolution that is important in measuring the iron
and iron-nickel line features at $\sim$6.7 and $\sim$8~keV,
respectively \citep{PhillipsK2004ApJ...605..921P},	
and the instrumental	
lines at $\sim$8 and $\sim$10~keV. We used 1~keV wide energy bins (three
native bins wide) at energies between 15 and
100~keV, where the highest resolution was not needed
to determine the parameters of the 	
continuum emission in this range.

We recommend the following sequential steps, which we generally followed, 
to obtain the ``best-fit" values of the spectral parameters and their
uncertainties in each time interval throughout the flare.

\begin{enumerate}

  \item Select a time interval that covers all of the \hsi
  observations for the flare of interest. 
  Also include times during the neighboring \hsi nighttime 	
  just before and/or just after the flare for use in
  determining the nonsolar background spectrum.



  \item Accumulate count-rate spectra corrected for live time, decimation, and pulse pile-up
  \citep[][although it is best to correct pile-up in step 6 below]{SmithD2002SoPh..210...33S}
  for each of the nine detectors in 4-s time bins (about one spacecraft spin period) for
  the full duration selected in step 1 above.
  A full response matrix, including off-diagonal elements, 
  is generated for each detector to relate the photon flux
  to the measured count rates in each energy bin.

  \item Import the count-rate spectrum and the corresponding
  response matrix for one of the detectors into 
  the \hsi spectral analysis routine, OSPEX. 
  We used detector 4 since it has close to the best energy resolution
  of all the detectors.

  \item Select time intervals to be used in estimating the
  background spectrum and its possible variation during the
  flare.  In general, nighttime data must be used if the attenuator state
  changes during the flare; otherwise pre- and/or post-flare
  spectra can be used. Account can be taken of orbital background
  variations during the flare by using a polynomial fit to
  the background time history in selectable energy ranges
  or by using the variations at energies above those
  influenced by the flare. For this event, since the thin attenuator
  was in place for the whole duration of the flare 
  a pre-flare interval was used for background estimation.

  \item Select multiples of the 4-s time intervals used in
  step 2 that are long enough to provide sufficient counts and 
  short enough to show the expected variations in the spectra as the flare
  progresses. Be sure that no time interval includes an
  attenuator change.  
  For this event, we selected the seven time intervals marked in Figure~\ref{lc.ps}
  covering the first HXR peak.

  \item {Fit the spectrum for the interval near the peak of
  the event to the desired functional form. 
  Spectra can be fitted to the algebraic sum of a variety of functional
  forms, ranging from simple isothermal and power-law functions to
  more sophisticated models, such as various multi-thermal models and thin- and thick-target
  models with a power-law electron spectrum having 
  sharp low-energy and high-energy cutoffs. 		
  In our case, we assumed an isothermal
  component plus a double power-law	
  to provides acceptable fits to the measured count-rate spectra in most cases.
  This simple two-component model is sufficient to
  capture the key physics for this flare, i.e., to estimate the
  relative contributions of
  the thermal and nonthermal components of the X-ray emission.

The isothermal spectrum was based on the predictions using the CHIANTI package 	
\citep[v. 5.2,][]{DereK1997A&AS.CHIANTI, YoungP2003ApJS.CHIANTI}  
in SSW with \citet{MazzottaP1998A&AS..133..403M}		
ionization balance.  The iron and nickel abundances were allowed to
vary about their coronal values to give the best fit to the iron features in the spectra.

  For simplicity, we set the power-law index below the variable break energy to be
fixed at $\gamma = 2$ to approximate a flat (constant)
electron flux below a cutoff energy.  The value of $\gamma$
above the break energy and the break energy itself were both treated as free parameters in
the fitting process.

We also included several other functions to accommodate
various instrumental effects. These included two narrow
Gaussians near 8 and 10~keV, respectively, to account for
two instrumental features that may be L-shell lines from
the tungsten grids. The thin attenuator was in place during
the entire course of the flare thus restricting us from
fitting the spectra below $\sim$6~keV.

  Another routine available in OSPEX was used to both offset the energy calibration and change
  the detector resolution to better fit the iron-line feature
  at $\sim$6.7~keV. This is important at high counting rates when 	
  the energy scale can change by up to $\sim$0.3~keV.

  Pulse pile-up can best be corrected for
  at this stage using a separate routine with 	
  count-rate dependent parameters although this is
  still in the developmental stage and was not used for this paper.
  However, the average live time (between data gaps) during the impulsive peak (interval 1, 08:20:27--08:20:56 UT) of 
  this M1.4 flare was 93.4\%.  This is to be compared with the values
  of 55\% and 94\% for the 2002 July 23 X4.8 flare and the 2002 February 20 C7.5 flare, respectively.
  In addition, the estimated ratio of piled-up counts to the total counts is below 10\% at all energies,
  indicating very minor pile-up effects on the spectra of this event.  A more detailed account
  on estimating pile-up severity can be found in \citet[][\S 2.1]{LiuW2006ApJ...649.1124L}
  and particularly for imaging spectroscopy in \citet{LiuW_FPAsym_2008ApJ}.


  It is important to use good starting values of the parameters
  to ensure that the minimization routine
  converges on the best-fit values. These were obtained 
  for detector 4 in the interval at the peak of the flare
  by experienced trial and error.}

  \item {Once an acceptable fit (reduced $\chi^2 \lesssim{2}$, 
  with the systematic uncertainties set to zero)
  is obtained to the spectrum for the peak interval,
  OSPEX has the capability
  to proceed either forwards or backwards in time to fit the
  count-rate spectra in other intervals using the best-fit parameters
  obtained for one interval as the starting parameters to
  fit the spectrum in
  the next interval. This reduces the time taken to fit each
  time interval but various manual adjustments are usually required
  to the fitted energy range, the required functions, etc.,
  in specific intervals to ensure adequate fits in each case with acceptable values of $\chi^2$.}

  \item The best-fit parameters found for each time
  interval for the one detector chosen in step 3 are now
  used as the starting parameters in OSPEX for the other
  detectors. In this way, acceptable fits can be obtained in
  each time interval for all nine detectors. In practice, it
  is usually not possible to include detectors 2, 5, or 7 in
  this automatic procedure since they have higher energy
  thresholds and/or poorer resolution compared to the other
  detectors.

  \item The different best-fit values (in practice, only
  six were obtained)	
  of each spectral parameter can now be combined to give a mean
  and standard deviation. These values then constitute the
  results of this spectral analysis and can be used for further
  interpretation as indicated in the body of the paper.
  For display purposes, it is important to show the best-fit photon spectrum computed
  using these mean parameters with some indication of the
  photon fluxes determined in each energy bin from the
  measured count rates. For this purpose, we have chosen to display 
  the photon fluxes averaged over over all detectors used in the analysis
  (all but detectors 2, 5, and 7).  The photon flux of each detector
  was determined by taking the count rate and folding it through the corresponding
  response matrix with the assumed photon spectrum having the best-fit parameters.
  This gives a reasonable representation but it
  is well known that data points determined in this way
  are ``obliging" and follow the assumed spectrum
  \citep{FenimoreE1983AdSpR...3..207F, FenimoreE1988ApJ...335L..71F}. 
  Hence, such plots (Figure~\ref{chp2LT_imgspec}) should be viewed with caution. 
  Also note that the $\chi^2$ values of the averaged photon fluxes are not necessarily representative of 
  the independent fits to the data of individual detectors.


\end{enumerate}

\subsection{Spatially Resolved Spectra}
\label{appendix_img-spec}

In order to determine the photon spectra of the two
distinct sources seen in the X-ray images, we used \hsi's
imaging spectroscopy capability and carried out the following steps.

\begin{enumerate}

 \item
We select the same seven intervals (marked in Figure~\ref{lc.ps}) 
as those used for the spatially integrated spectra.

 \item
For each selected time interval, images in narrow energy bins ranging from
1~keV wide at 6~keV to 11~keV wide at 50~keV 
were constructed using the computationally expensive PIXON algorithm 
\citep{MetcalfT1996ApJ...466..585M}, which gives the best photometry
and spatial resolution \citep{AschwandenMJ2004SoPh.HESSI.img.soft}
among the currently available imaging algorithms.
Detectors 3, 4, 5, 6, and 8 covering angular
scales between $6 \farcs 8$ and $106\arcsec$ were used to allow the two
sources	
to be clearly resolved. 
No modulation was evident in the detector 1 and 2 count rates, showing that
the sources had no structure finer than the $3 \farcs 9$ FWHM resolution of detector 2.

 \item
The PIXON images were imported into OSPEX for extracting fluxes of individual sources.
Note that the images are provided in units of 
[photons~$\unit{cm}^{-2} \unit{s}^{-1} \unit{keV}^{-1} \unit{arcsec}^{-2}$]
using only the {\it diagonal}
elements of the detector response matrix to convert from the
measured count rates to photon fluxes. OSPEX converts the
images back to units of 
[counts~$\unit{cm}^{-2} \unit{s}^{-1} \unit{keV}^{-1} \unit{arcsec}^{-2}$]
using the same diagonal elements and then uses the {\it full}
detector response matrix, including all off-diagonal
elements, to compute the best-fit photon spectrum 
(photons~$\unit{cm}^{-2} \unit{s}^{-1} \unit{keV}^{-1}$)
for each source separately. The summed count rates in the two boxes shown
in the middle panel of Figure~\ref{hsi_multi.ps} around the
average positions of the two sources were accumulated
separately for each image in each energy bin. The boxes were adjusted 
accordingly for each time interval if the sources moved. 
(Note that only a single		
box was used for Interval~1 when only the lower source was detected.) 

 \item
 The uncertainties in the count rates were calculated from the PIXON
error map based on $\chi^2$ variations of the reconstructed image \citep[see][\S A.2]{LiuW2006PhDT........35L}. 
The errors were originally obtained in photon space
and then converted in the same way described above to count space where
the actual fitting was performed.

 \item
The two independent count-rate spectra,
one for each source, were then fitted independently to the same
functions used for the spatially integrated spectra as
described earlier.  We further demand that the iron abundance of the thermal
component and the break energy of the double power-law are fixed at the
values given by the fit to the corresponding spatially integrated spectrum 
in the same time interval. This makes the spectra directly comparable for
our purposes.
Note that the error bars of the imaging spectral parameters are obtained from the $\chi^2$ variation during the 
fitting procedure.  At times when such an error is smaller than that of the corresponding spatially integrated
spectrum, the latter value is used instead.

\end{enumerate}

  Finally, for a self-consistency check, we have
compared the sum of the imaging spectra of the two sources with the spatially integrated spectrum
and found they are consistent.	
The only exception is at the low energies
($\lesssim$10~keV) where the imaging spectra do not have enough resolution 	
to see the iron-line feature.



\end{document}